\documentclass[12pt]{article}
\usepackage{amsfonts,amsmath,amssymb,epsfig,here,latexsym,lscape,setspace,xfrac}
\usepackage[english]{babel}
\usepackage{hyperref}
\graphicspath{{c:/swp55/docs/}}

\newcommand{\be}{\begin{equation}}
\newcommand{\ee}{\end{equation}}
\newcommand{\bea}{\begin{eqnarray}}
\newcommand{\eea}{\end{eqnarray}}
\newcommand{\beas}{\begin{eqnarray*}}
\newcommand{\eeas}{\end{eqnarray*}}

\newtheorem{theorem}{Theorem}[section]

\newtheorem{proposition}[theorem]{Proposition}

\newtheorem{remark}[theorem]{Remark}
\newtheorem{example}[theorem]{Example}
\newtheorem{examples}[theorem]{Examples}
\newtheorem{foo}[theorem]{Remarks}

\newenvironment{Remark}{\begin{remark}\rm}{\end{remark}}

\newenvironment{proof}{\addvspace{\medskipamount}\par\noindent{\it Proof}}
{\unskip\nobreak\hfill$\Box$\par\addvspace{\medskipamount}}

\newcommand{\E}[1]{{\rm E}\left[#1\right]}
\newcommand{\pr}[1]{{\rm P}\left[#1\right]}

\DeclareMathOperator{\argmax}{arg\,max}

\begin{document}
\title{\vskip -1cm Risk Apportionment: The Dual Story\thanks{
With great sadness, we lost our friend and co-author Harris Schlesinger,
who passed away while we were in the process of writing this paper.
We are very grateful for detailed comments from
Sebastian Ebert, Johanna Etner, Christian Gollier, Glenn Harrison, Mike Hoy, Liqun Liu (discussant), Lisa Posey (discussant), Nicolas Treich, Michel Vellekoop, and Claudio Zoli.
We are also grateful to
conference and seminar participants
at the EGRIE 
Meeting, 
the Risk Theory Society 
Meeting, 
the World Risk and Insurance Economics Congress, 
the Tinbergen Institute,
and the KAFEE seminar at the University of Amsterdam for helpful comments.
This paper was circulated earlier under the title ``Prudence, temperance (and other virtues): The dual story''.
This research was funded in part by the Netherlands Organization for
Scientific Research (Laeven) under grant NWO VIDI.}}
\author{Louis R. Eeckhoudt\\
{\footnotesize IESEG School of Management}\\
{\footnotesize CNRS-LEM UMR 9221}\\
{\footnotesize and CORE}\\
{\footnotesize {\tt Louis.Eeckhoudt@fucam.ac.be}}\\\and Roger J. A. Laeven\\
{\footnotesize Amsterdam School of Economics}\\
{\footnotesize University of Amsterdam, EURANDOM}\\
{\footnotesize and CentER}\\
{\footnotesize {\tt R.J.A.Laeven@uva.nl}}\\\and Harris Schlesinger\\
{\footnotesize Department of Finance}\\
{\footnotesize University of Alabama}\\
{\footnotesize and University of Konstanz}\\
{\footnotesize {\tt hschlesi@cba.ua.edu}}}
\date{This Version:
\today} \maketitle
\begin{abstract}
\noindent
By specifying model free preferences towards simple nested classes of lottery pairs,
we develop the dual story to stand on equal footing with that of
(primal) risk apportionment.
The dual story provides an intuitive interpretation, and full characterization,
of dual counterparts of such concepts as prudence and temperance.
The direction of preference between these nested classes of lottery pairs
is equivalent to signing the successive derivatives of the probability weighting function
within Yaari's (1987) dual theory.
We explore implications of our results for optimal portfolio choice
and show that the sign of the third 
derivative of the probability weighting function
may be naturally linked to a self-protection problem.
\\[1mm]\noindent\textbf{Keywords:} Higher Order Risk Attitudes; 
Prudence; Temperance;
Risk Apportionment; Dual Theory; 
Portfolio Choice; Self-Protection.
\\[1mm]\noindent\textbf{JEL Classification:} D81, G11, G22.
\end{abstract}

\makeatletter
\makeatother
\maketitle

\newpage

%

\section{Introduction}


Although 
first received with some skepticism,
the notions of prudence and temperance have now been widely accepted
almost on par with the fundamental concept of risk aversion,
at least in an expected utility (EU) framework.

The expanding use of these notions,
sometimes termed ``higher order risk attitudes'',
can be explained by the fact that they were progressively given
a more general interpretation. 
Consider prudence, for instance.
This term was coined by Kimball (1990)
in an influential paper in which he showed that precautionary savings as an optimizing type of behavior
is characterized in an EU
framework 
by a positive third derivative of the utility function (i.e., ``$U'''\geq 0$'' or ``prudence'').
However, it is by now well-known that this positive sign of 
$U'''$ can be justified more generally
outside the specific decision problem of saving.
This more primitive justification of prudence was initiated by Menezes, Geiss and Tressler (1980),
who used the term ``downside risk aversion'', and it was further pursued in Eeckhoudt and Schlesinger (2006),
who also showed how to proceed from prudence to higher order risk attitudes.
These authors first state a ``model free'' preference,
namely that decision makers (DM's) like to ``combine good with bad''
instead of having to face either everything good or everything bad.
Next, this model free preference is shown to be translated into prudence ($U'''\geq 0$) within the EU model,
and from prudence---by defining a sequence of nested lotteries
and always asserting the preference for combining good with bad---the higher order risk attitudes may be obtained similarly,
starting with temperance ($U''''\leq 0$) at the fourth order.

It turns out besides that the simple primitive interpretation of prudence and higher order risk attitudes
found in Eeckhoudt and Schlesinger (2006) lends itself easily to experimental verification.
As a result, there is now an intensive experimental research activity around the concepts of
prudence and temperance in an EU framework (e.g., Ebert and Wiesen (2011, 2014),
Deck and Schlesinger (2010, 2014), and Noussair, Trautmann and van de Kuilen (2014),
to name a few).

The preference for ``combining good with bad'' has appeared under different names in the literature.
It was called ``risk apportionment'' in Eeckhoudt and Schlesinger (2006).
A little earlier, Chiu (2005) referred to a ``precedence relation'',
in which one stochastic dominant change 
precedes another. 
The phrase ``combining good with bad'' as a primitive trait first appeared in
Eeckhoudt, Schlesinger and Tsetlin (2009).

While prudence, temperance, and higher order risk attitudes can be presented initially as
natural properties in a model free environment, their interpretation and implementation have been developed so far exclusively
within an EU framework.\footnote{In very recent work, Baillon (2017)
generalizes these interpretations of prudence and higher order risk attitudes to a setting featuring ambiguity.}
In this paper, by specifying new model free preferences towards simple nested classes of lottery pairs,
we develop the dual story to stand on equal footing with that of
(primal) risk apportionment.
The dual story provides an intuitive interpretation, and full characterization,
of dual counterparts of such concepts as prudence, temperance, and other virtues.
We show that the direction of preference between the nested classes of lottery pairs that we construct
is equivalent to signing the $m^{\mathrm{th}}$ derivative of the probability weighting (or distortion) function
within Yaari's (1979) dual theory (DT), with $m$ an arbitrary positive integer.\footnote{One may therefore say that
this paper constitutes the genuine appropriate dual counterpart of (primal) risk apportionment (Eeckhoudt and Schlesinger (2006)).
Indeed, from a technical perspective, the contribution of the present paper to the extant literature on (dual or inverse) stochastic ordering in DT
(e.g., Muliere and Scarsini (1989), Wang and Young (1998) and Chateauneuf, Gajdos and Wilthien (2002))
is similar to the contribution of Eeckhoudt and Schlesinger (2006) to
the literature on (primal) stochastic ordering in EU
(e.g., Whitmore (1970), Ekern (1980) and Menezes, Geiss and Tressler (1980)).}
It turns out that this development requires a fundamental departure from the approach of Eeckhoudt and Schlesinger (2006),
which, as we will show, is unable to deliver the desired implications within the DT framework.
The dual story we develop retains generic features of the primal story---e.g., a precedence relation---but crucially departs from it in its 
construction and implementation---e.g., by reference to what we will refer to as ``squeezing'' and ``anti-squeezing'' and to the ``dual moments''.

This paper thus represents 
a first step towards a more general interpretation of higher order risk attitudes within alternative non-EU decision models, 
such as rank-dependent utility and prospect theory
(Kahneman and Tversky (1979), Quiggin (1982), Schmeidler (1986, 1989), Tversky and Kahneman (1992)),
for which DT is a building block.
Indeed, because DT is ``orthogonal'' to EU,
our analysis not only reveals the differences with the primal story under EU,
but it is also a prerequisite for a 
development of higher order risk attitudes as primitive traits of behavior
compatible with the more general models of choice under risk
provided by rank-dependent utility and prospect theory.

A positive sign of the third derivative of the probability weighting function is consistent
with an ``inverse S-shape'',
exhibited by the popular probability weighting functions proposed by Tversky and Kahneman (1992)
(see also Wu and Gonzalez (1996, 1998)) and Prelec (1998) under typical parameter sets implied by experiments.\footnote{The inverse S-shape
reflects the psychological notion of diminishing sensitivity (in the domain of probabilities), which stipulates that DM's become less (more) sensitive to changes in the objective probabilities when they move away from (towards) the reference points 0 and 1.}
These inverse S-shaped probability weighting functions typically feature a positive sign for the odd derivatives
and an alternating sign (first negative at low wealth levels, then positive at high wealth levels) for the even derivatives.
Provided that
the probability weighting function is first concave and then convex with second derivative equal to zero at the inflection point,
a positive sign of the third derivative of the probability weighting function
implies that the function becomes more concave when moving to the left of the inflection point
and becomes more convex when moving to the right of the inflection point.

As is well-known, primal and dual stochastic dominance coincide up to the second order
and may diverge from the third order onwards.
As a by-product, which is of interest in its own right,
the model free story appropriate for DT will nicely make apparent the fundamental reason
behind this divergence.

Our results about the shape of the probability weighting function are relevant for the analysis of well-known problems in the economics of risk,
such as portfolio choice and the level of self-protection in the presence of background risk.
We first illustrate our results by deriving their implications for optimal portfolio choice with a risky asset,
a risk-free asset and access to zero-mean financial derivative products on the risky asset.
We show that, contrary to under EU (Gollier (1995)), an $m^{\mathrm{th}}$ order improvement of the risky asset's return,
achieved by supplementing (hence, squeezing) the risky asset with an appropriate selection of derivative products (e.g., a straddle at the third order
or a volatility spread at the fourth order), never reduces the demand for the risky asset under the DT model.
Furthermore, we show that the third derivative of the probability weighting function naturally appears
in a self-protection problem that trades off the risk of a loss and the effort of protecting against the loss,
in the presence of an independent background risk.
In particular, if 
the third derivative of the probability weighting function is positive (``dual prudence''),
the background risk stimulates self-protection.

This paper is organized as follows.
In Section \ref{Sec:Prel}, we fix the notation and setting,
introduce some preliminaries for the DT decision model, 
and provide some basic intuition behind our results.
In Section \ref{Sec:Pref}, we introduce dual higher order risk attitudes
by developing new model free preferences. 
Section \ref{Sec:PC} illustrates implications of our results for optimal portfolio choice.
Section \ref{Sec:Prot} shows that the sign of the third derivative of the probability weighting function
is naturally linked to a self-protection problem.
Section \ref{Sec:Res} contains the formal presentation of our general results.
We conclude in Section \ref{Sec:Con} with a summary of the results
and an indication of potential extensions.
Proofs are relegated to the Appendix.\footnote{Some supplementary material,
including two 
illustrations and some technical details to supplement Section \ref{Sec:Int},
some further details to supplement Section \ref{Sec:Prot},
and some technical details to supplement the Appendix, 
suppressed in this version to save space, is contained
in the extended online version of this paper that is available from the authors' webpages (see {\tt http://www.rogerlaeven.com}).}



\section{Preliminaries}\label{Sec:Prel}


\subsection{Notation and Setting}

We represent an $n$-state lottery $A$, assigning probabilities $p_{i}$ to final wealth outcomes $x_{i}$, $i=1,\ldots,n$,
by $A=[\ x_1,p_1\ ;\ \ldots\ ;\ x_n,p_n\ ]$.
We always assume that states are ordered according to their associated outcomes,
from the lowest outcome state to the highest outcome state.
Outcomes are assumed to be non-negative.
That is,
$0\leq x_{1}\leq\cdots\leq x_{n}$.
With each lottery, generating a probability distribution over outcomes, one can associate a random variable.
Henceforth, the lottery and its associated random variable are often identified.
We write $\pr{A\leq x}=F_{A}(x)=1-S_{A}(x)$,
with $F_{A}$ and $S_{A}$ the cumulative and decumulative distribution functions of $A$, respectively.
Furthermore, we denote by $\succeq$ a (weak) preference relation over lotteries.

Under Yaari's (1987) dual theory (DT), the evaluation 
$V$ of the $n$-state lottery $A$ is given by
\begin{align}
V[A]&=\int_{0}^{\infty}x\,\mathrm{d}h(F_{A}(x))\nonumber\\
&=\sum_{i=1}^{n}x_{i}\left(h(F_{A}(x_{i}))-h(F_{A}(x_{i-1}))\right),
\label{eq:V}
\end{align}
with $x_{0}=0$ and $F_{A}(x_{0})=0$ by convention,
and with $h:[0,1]\rightarrow[0,1]$ satisfying $h(0)=0$, $h(1)=1$, and $h'\geq 0$, a probability weighting (or distortion) function,
henceforth assumed 
to be differentiable for all degrees of differentiation on $(0,1)$.\footnote{In the literature, following Yaari (1987),
the evaluation under DT is often carried out by distorting the decumulative distribution function $S_{A}$
instead of the cumulative distribution function $F_{A}$ as in \eqref{eq:V}.
Notice, however, that
\begin{align*}V[A]&=\int_{0}^{\infty}x\,\mathrm{d}h(F_{A}(x))=\int_{0}^{\infty}x\,\mathrm{d}\left(1-\bar{h}(S_{A}(x))\right)
=\int_{0}^{\infty}\bar{h}(S_{A}(x))\,\mathrm{d}x
,\end{align*}
with $\bar{h}(p):=1-h(1-p)$,
mapping the unit interval onto itself and satisfying $\bar{h}(0)=0$, $\bar{h}(1)=1$, and $\bar{h}'\geq 0$,
and that positive signs of the higher order derivatives of $\bar{h}$ are equivalent to alternating signs of the higher order derivatives of $h$.
In particular, concavity of $h$ translates to convexity of $\bar{h}$.
To be consistent with the
alternating signs of the derivatives of the utility function in the EU model,
we define the evaluation $V$ under DT by distorting the cumulative distribution function $F_{A}$
rather than the decumulative distribution function $S_{A}$.
Thus, in our setting a concave probability weighting function $h$
is equivalent to ``strong'' risk aversion in the sense of aversion to mean preserving spreads
(Chew, Karni and Safra (1987) and Ro\"ell (1987)), and,
more generally, the higher order derivatives of the probability weighting function $h$ will naturally alternate in sign, just like that of $U$.
}
We sometimes denote by $h^{(m)}$ the $m^{\mathrm{th}}$ derivative of $h$.\footnote{We use the notations $h'$, $h'',\dots$ and $h^{(1)}$, $h^{(2)},\ldots$ interchangeably.}

We say that a lottery $B$ dominates a lottery $A$ 
in third-degree dual (or inverse) stochastic dominance order if
\begin{equation*}\E{A}\leq\E{B},\qquad\E{\min(A_1,A_2)}\leq\E{\min(B_1,B_2)},\quad\mathrm{and}\quad ^{3}F_{A}^{-1}\leq\ ^{3}F_{B}^{-1}.\end{equation*}
Here, $A_1$ ($B_1$) and $A_2$ ($B_2$) are two independent draws from lottery $A$ ($B$).
Furthermore,
$^{m+1}F_{A}^{-1}(q)=\int_{0}^{q}\ ^{m}F_{A}^{-1}(p)\,\mathrm{d}p$, $m=1,2,\ldots$, $0\leq q\leq 1$,
with $^{1}F_{A}^{-1}\equiv F_{A}^{-1}$,
and inequalities between functions are understood pointwise.
We refer to $\E{\min(A_1,A_2,\ldots,A_{m})}$ as the $m^{\mathrm{th}}$ ``dual moment'' of $A$.\footnote{In statistics,
these moments are sometimes referred to as mean (first) order statistics.
They measure the expected worst outcome in an experiment with repeated independent draws; see also David (1981).
Primal moments occur under EU when considering power functions as utility functions.
Similarly, dual moments occur under DT when considering power functions as probability weighting functions:
\begin{align*}
\E{\min(A_1,A_2,\ldots,A_{m})}&=\int_{0}^{\infty}(1-F_{A}(x))^{m}\,\mathrm{d}x
=\int_{0}^{\infty}\bar{h}(1-F_{A}(x))\,\mathrm{d}x
=\int_{0}^{\infty}x\,\mathrm{d}h(F_{A}(x))\\
&=V[A],\end{align*}
with $\bar{h}(p)=p^{m}$ and $\bar{h}(p)=1-h(1-p)$, $0\leq p \leq 1$.
Observe that the power probability weighting function satisfies $\bar{h}''\geq 0$, hence $h''\leq 0$.
}
In full generality, we say that lottery $B$ dominates lottery $A$ 
in $m^{\mathrm{th}}$-degree dual stochastic dominance order ($m=2,3,\ldots$) if
\begin{align*}\E{A}&\leq\E{B},\quad\E{\min(A_1,A_2)}\leq\E{\min(B_1,B_2)},\quad \ldots\quad ,\\
\quad\E{\min(A_1,A_2,\ldots,A_{m-1})}&\leq\E{\min(B_1,B_2,\ldots,B_{m-1})},\quad\mathrm{and}\quad ^{m}F_{A}^{-1}\leq\ ^{m}F_{B}^{-1}.\end{align*}
Preferences for $m^{\mathrm{th}}$ order dual stochastic dominance, with first $(m-1)$ dual moments equal, are well-known to be linked
to the sign of the $m^{\mathrm{th}}$ derivative of the probability weighting function within DT (Muliere and Scarsini (1989)),
just like primal stochastic dominance orders are connected to the successive derivatives of the utility function under EU (Ekern (1980)).
For further details on dual (inverse) stochastic dominance, we refer e.g., to De La Cal and C\'arcamo (2010) and the references therein.

Under EU the sign of the $m^{\mathrm{th}}$ derivative of the utility function
can be interpreted by comparing simple nested pairs of lotteries with equal $(m-1)$ primal moments,
obtained via the apportionment of harms (Eeckhoudt and Schlesinger (2006)).
To interpret the sign of the $m^{\mathrm{th}}$ derivative of the probability weighting function under DT,
we develop simple nested lottery pairs that have equal $(m-1)$ dual moments.

\subsection{Illustration and Intuition}\label{Sec:Int}

Since it is well-known that EU and DT agree
at the first and second orders
in their evaluation of a sure reduction in wealth and a mean preserving spread
(see Yaari (1986), Chew, Karni and Safra (1987), Ro\"ell (1987)
and Muliere and Scarsini (1989)) 
we develop here a numerical example to illustrate that they may diverge at the third order.
This example builds the intuition behind the reason why EU and DT may diverge at the third order
and motivates the adjustments that have to be made to the story of ``combining good with bad''
used to interpret the signs of the successive derivatives of the utility function under EU.
While it is known that
third order primal and dual stochastic dominance do not in general coincide
(see Muliere and Scarsini (1989) and its references),
we illustrate their potential divergence here in the context of (primal) risk apportionment
(Eeckhoudt and Schlesinger (2006)).\footnote{In this context, similarly,
EU and DT may agree at the third order,
such as in their evaluation of the specific case of Mao's (1970) lotteries considered also by Menezes, Geiss and Tressler (1980)
(see the extended online version),
or may not agree, such as in the example presented in this section.}

We start from an initial lottery $I$ given by
\begin{equation*}
I=[\ 2,1/2\ ;\ 3,1/2\ ],
\end{equation*}
and, invoking (primal) risk apportionment, ``add'' the independent zero-mean risk $\tilde{\varepsilon}$ given by
\begin{equation*}
\tilde{\varepsilon}=[\ -2,1/3\ ;\ 1,2/3\ ],
\end{equation*}
to one of the states of $I$.
If $\tilde{\varepsilon}$ is added to the first state of $I$ (which is the bad one)
we generate a lottery $A$ given by
\begin{equation*}A=[\ 0,1/6\ ;\ 3,5/6\ ],\end{equation*}
while if it is added to the second state (which is the good one)
we generate a lottery $B$ given by
\begin{equation*}B=[\ 1,1/6\ ;\ 2,1/2\ ;\ 4,1/3\ ].\end{equation*}
From Eeckhoudt and Schlesinger (2006), we know that under EU,
$U'''\geq 0$ implies $B\succeq A$.\footnote{Lottery $B$ is obtained from lottery $I$
by adding the bad risk $\tilde{\varepsilon}$ (bad since $\tilde{\varepsilon}$ is second order stochastically dominated by $0$)
to the good state of $I$ while the converse is true for $A$.
One might alternatively say
that in $A$ the bad ($\tilde{\varepsilon}$) precedes the good ($0$).
Conversely, in $B$ the good ($0$) precedes the bad ($\tilde{\varepsilon}$).}
If the DM has a quadratic utility (hence, ``$U'''=0$'' or ``zero prudence'')
he is indifferent between $A$ and $B$.

Now consider a DM under the DT model with a quadratic probability weighting function:
\begin{equation*}
h(p)=\alpha p - \beta p^2,
\end{equation*}
with $\alpha=1+\beta$ and $0\leq\beta\leq 1$ so that
$h(0)=0$, $h(1)=1$ and $h$ is non-decreasing and concave.\footnote{Note that $h(0)=0$ by definition.
Furthermore, to have $h(1)=1$, $\alpha-\beta=1$ should hold.
Hence, $h(p)=(1+\beta)p-\beta p^2$, so that $h'(p)=1+\beta-2\beta p$.
To have $h'(1)\geq 0$ (hence $h'(p)\geq 0$ whenever $h''(p)\leq 0$), $\beta\leq 1$ should hold.
Finally, $h''(p)=-2\beta$ so $h''\leq 0$ if $\beta\geq 0$.
In sum, $\alpha=1+\beta$ and $0\leq\beta\leq 1$.}
This DM has ``zero dual prudence'' (``$h'''=0$'').
Simple computations reveal that for this DM, $V[A]\geq V[B]$
with strict inequality whenever $\beta>0$ (hence, $h''<0$).
Thus, there is a difference of opinions between the EU DM, who is indifferent between $A$ and $B$
under zero primal prudence,
and the DT DM, who prefers $A$ over $B$ under zero dual prudence.

The divergence of opinions between these two DM's
can be related to a difference between the primal and dual moments of $A$ and $B$.
Clearly, $A$ and $B$ have the first two primal moments (mean and variance) in common.
Their second dual moments are, however, different.\footnote{Indeed,
$\E{\min(A_1,A_2)}= 25/12$ while 
$\E{\min(B_1,B_2)}= 23/12$.
}

It thus appears that the model free prescription that favors ``combining good with bad''
in the particular way as suggested by Eeckhoudt and Schlesinger (2006)
leads to an interpretation of the signs of the successive derivatives of $U$ within the EU model;
however, this model free principle does not always generate a similar interpretation for the signs of the
successive derivatives of the probability weighting function under the DT framework.
Thus, if one desires to obtain for DT and the signs of the successive derivatives of the probability weighting function
a development that parallels that within EU, one has to modify the initial model free preferences.
This is the purpose of the next section.

\setcounter{equation}{0}

\section{Model Free Preferences} 
\label{Sec:Pref}


To provide an intuitive interpretation to the signs of the successive derivatives
of the probability weighting function beyond the second order\footnote{As is well-known, 
both $U''\leq 0$ and $h''\leq 0$ correspond to strong risk aversion
(i.e., aversion to mean preserving spreads); see the references in Section \ref{Sec:Int}.}
we now develop the appropriate model free preferences. 
We will continue to assert that DM's
want to combine good with bad.
In Chiu's (2005) equivalent terminology,
the DM still satisfies a precedence relation.
He favors that the good precedes the bad.
However, our definition of good and bad will be based on the concepts of ``squeezing'' and ``anti-squeezing'' a distribution
instead of adding zero-mean risks and zero with certainty as in Eeckhoudt and Schlesinger (2006).
Squeezing and anti-squeezing will serve as the main building blocks in our approach.
The sequence of squeezes and anti-squeezes that we develop will preserve dual rather than primal moments.

Squeezing occurs when we transform an initial lottery
\begin{equation*}
L=[\ x_1,p_1\ ;\ \ldots\ ;\ x_{i},p_{i}+p\ ;\ \ldots\ ;\ x_{j},p_{j}+p\ ;\ \ldots\ ;\ x_{n},p_{n}\ ],
\end{equation*}
with $p_{k}\geq 0$ (where equality to zero is explicitly permitted), $k=1,\ldots,n$,
and $p>0$,
into a lottery $D$ given by
\begin{equation*}
D=[\ x_1,p_1\ ;\ \ldots\ ;\ x_{i},p_{i}\ ;\ x_{i}+x,p\ ;\ \ldots\ ;\ x_{j}-x,p\ ;\ x_{j},p_{j}\ ;\ \ldots\ ;\ x_{n},p_{n}\ ],
\end{equation*}
with $x>0$.
Anti-squeezing occurs when $x$ is replaced by $-x$, that is, when $L$ is transformed into a lottery $C$ given by
\begin{equation*}
C=[\ x_1,p_1\ ;\ \ldots\ ;\ x_{i}-x,p_{i}\ ;\ x_{i},p\ ;\ \ldots\ ;\ x_{j},p\ ;\ x_{j}+x,p_{j}\ ;\ \ldots\ ;\ x_{n},p_{n}\ ].
\end{equation*}
Clearly, both squeezing and anti-squeezing preserve the mean.
Note the generality of these transformations.
Depending on the specification of $x$ and $p$,
a squeeze and an anti-squeeze can be interpreted both as a shift in outcomes and as a shift in probabilities.

\begin{figure}[t]
\begin{center}
\caption{Squeezing and Anti-Squeezing at the Second Order
\newline\footnotesize This figure plots the transformation from $L^{(2)}$ to $D^{(2)}$ (left panel) and from $L^{(2)}$ to $C^{(2)}$ (right panel),
with $x_{1}=1$, $x_{2}=2$, and $p=1/2$.
$\mathcal{G}^{(3)}$ and $\mathcal{B}^{(3)}$ will be used to develop the dual story at the third order.
}
\vskip 0.4 cm
\includegraphics[scale=1.22,angle=0]{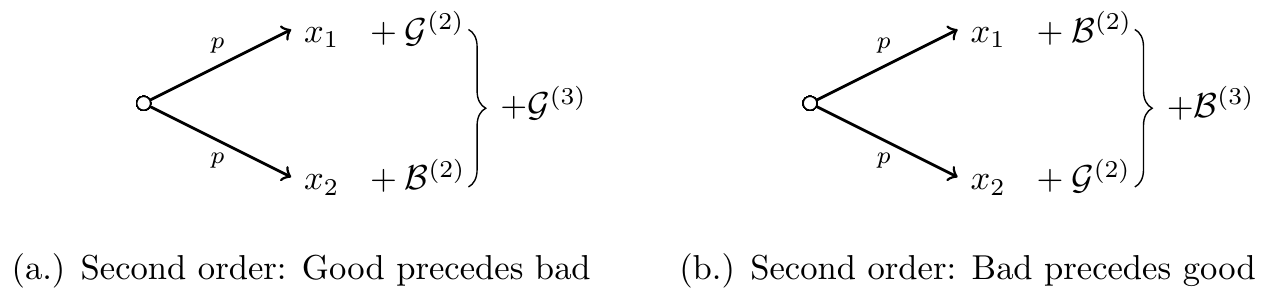}
\label{fig:2nd}
\end{center}
\end{figure}

In this section, we explicate the dual story in examples.
Section \ref{Sec:Res} contains the general approach.
We first return to the second order and consider an initial lottery $L^{(2)}$ given by\footnote{The superscript $^{(2)}$ refers
to ``second order'' and more generally the superscript $^{(m)}$ refers to ``$m^{\mathrm{th}}$ order''.}
\begin{equation*}
L^{(2)}=[\ 1,1/2\ ;\ 2,1/2\ ].
\end{equation*}
Now we transform the lottery $L^{(2)}$ into a lottery $D^{(2)}$ by squeezing.
Specifically, we ``attach'' (state-wise with the state probabilities matched)
\begin{equation*}
\mathcal{G}^{(2)}=[\ x,p\ ]\quad\mathrm{and}\quad \mathcal{B}^{(2)}=[\ -x,p\ ],
\end{equation*}
to $L^{(2)}$,
where in this case we let $p=1/n$, $n=2$, and where $x=1/M$, $M\geq 2$.\footnote{The condition on $M$
guarantees that the squeeze does not change the initial ranking of outcomes.
}
We assume the good ($\mathcal{G}^{(2)}$) 
precedes the bad ($\mathcal{B}^{(2)}$).
This squeeze 
yields the new lottery $D^{(2)}$ given by
\begin{equation*}D^{(2)}=[\ 1+1/M,1/2\ ;\ 2-1/M,1/2\ ].\end{equation*}
Of course, an anti-squeeze of $L^{(2)}$ that is same-sized but opposite to the squeeze,
achieved by attaching $\mathcal{B}^{(2)}$ and $\mathcal{G}^{(2)}$  state-wise
with the bad now preceding the good,
would generate $C^{(2)}$
given by
\begin{equation*}C^{(2)}=[\ 1-1/M,1/2\ ;\ 2+1/M,1/2\ ].\end{equation*}
See the illustration in Figure \ref{fig:2nd}.

Clearly, under strong risk aversion, $D^{(2)}\succeq C^{(2)}$.\footnote{For reasons that become apparent in Section \ref{Sec:Res}
we don't compare the changed lotteries $C^{(2)}$ and $D^{(2)}$ to the initial lottery $L^{(2)}$ they are generated from,
although this comparison is straightforward in this particular case, in which $D^{(2)}\succeq L^{(2)}\succeq C^{(2)}$.
In fact, throughout this section, $D^{(m)}\succeq L^{(m)}\succeq C^{(m)}$, $m=2,3,4$, for DT DM's having probability weighting functions
with higher order derivatives that alternate in sign.
For brevity, we sometimes directly construct $D^{(m)}$ from $C^{(m)}$ in the following sections.}
These preferences correspond to 
$h''\leq 0$ under DT and to $U''\leq 0$ under the EU model.
Indeed, squeezing and anti-squeezing are special cases of a mean preserving contraction and a mean preserving spread
in the sense of Rothschild and Stiglitz (1970),
and DT and EU are known to agree
in their evaluation of a mean preserving contraction and a mean preserving spread.

Turning now to the third order, the agreement between DT and EU may collapse, as already anticipated in Section \ref{Sec:Int}.
The model free preferences based on the precedence relation towards good and bad coupled with the notions of squeezing and anti-squeezing are
deployed as follows.\footnote{This is again an illustration. The general treatment is contained in Section \ref{Sec:Res}.}
We start from an initial lottery $L^{(3)}$ given by
\begin{equation*}L^{(3)}=[\ 1,1/3\ ;\ 2,1/3\ ;\ 4,1/3\ ].\end{equation*}
Then we generate $D^{(3)}$ as
\begin{equation*}D^{(3)}=[\ 1+1/M,1/3\ ;\ 2-2/M,1/3\ ;\ 4+1/M,1/3\ ],\end{equation*}
with $M\geq 3$.
To obtain $D^{(3)}$ from $L^{(3)}$, we first transform the worst two states of $L^{(3)}$ 
by applying a squeeze, 
attaching
\begin{equation*}
\mathcal{G}^{(3)}=[\ x,p\ ;\ -x,p\ ],
\end{equation*}
where we take in this case $p=1/n$, $n=3$, and $x=1/M$, $M\geq 3$.
Then on the best two states of $L^{(3)}$ we perform an anti-squeeze,
achieved by attaching
\begin{equation*}
\mathcal{B}^{(3)}=[\ -x,p\ ;\ x,p\ ],
\end{equation*}
with $p=1/n$, $n=3$, and $x=1/M$, $M\geq 3$.
In the transformation from $L^{(3)}$ to $D^{(3)}$, the good ($\mathcal{G}^{(3)}$) precedes the bad ($\mathcal{B}^{(3)}$).
To keep the required number of states as small as possible,
for ease and parsimony of exposition,
we consider in this illustration a lottery $L^{(3)}$ with only 3 states,
having moreover equal probabilities of occurrence,
and we squeeze and anti-squeeze at an overlapping state (that with outcome 2) and by the same amount.
These restrictions are, however, not required for our general approach; see Section \ref{Sec:Res} for further details.

\begin{figure}[t]
\begin{center}
\caption{Squeezing and Anti-Squeezing Sequences at the Third Order
\newline\footnotesize This figure plots the transformation from $L^{(3)}$ to $D^{(3)}$ (left panel) and from $L^{(3)}$ to $C^{(3)}$ (right panel),
with $x_{1}=1$, $x_{2}=2$, $x_{3}=4$, and $p=1/3$.
$\mathcal{G}^{(4)}$ and $\mathcal{B}^{(4)}$ will be used to develop the dual story at the fourth order.
}
\vskip 0.4 cm
\includegraphics[scale=1.22,angle=0]{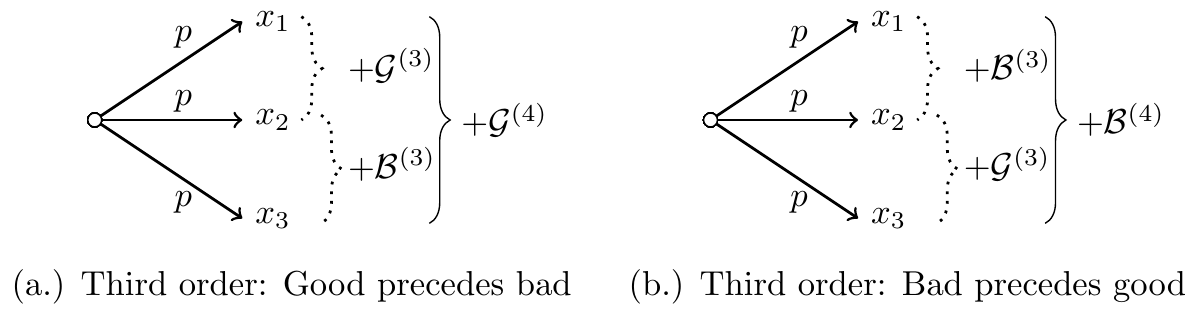}
\label{fig:3rd}
\end{center}
\end{figure}

Similarly, we generate a lottery $C^{(3)}$
given by
\begin{equation*}C^{(3)}=[\ 1-1/M,1/3\ ;\ 2+2/M,1/3\ ;\ 4-1/M,1/3\ ],\end{equation*}
obtained from the initial lottery $L^{(3)}$
by letting the bad, $\mathcal{B}^{(3)}$, precede the good, $\mathcal{G}^{(3)}$.
See the illustration in Figure \ref{fig:3rd}.

As shown in Section \ref{Sec:Res},
to which a precise statement of this result is deferred---see Theorems \ref{Th:Char3-1} and \ref{Th:Char3-2}---,
``dual prudence'' (or ``$h'''\geq\ 0$'') corresponds to ``$D^{(3)}\succeq C^{(3)}$''.
Both the well-known probability weighting function of Tversky and Kahneman (1992) and that of Prelec (1998)
exhibit $h'''\geq 0$ under the typical parameter sets implied by experiments.

It is very important to realize that
there is no unanimity among EU maximizers with $U'''\geq 0$
on the comparison between $C^{(3)}$ and $D^{(3)}$.
This is essentially so because
these two lotteries with the same mean have different variances and skewnesses.
Indeed, $\mathrm{Var}[D^{(3)}]>\mathrm{Var}[C^{(3)}]$ and $\mathrm{Skew}[D^{(3)}]>\mathrm{Skew}[C^{(3)}]$.
As a result, EU DM's who are relatively more (less) risk averse than prudent prefer $C^{(3)}$ ($D^{(3)}$).
While our story of squeezing and anti-squeezing may yield different primal moments (variances in particular) for the lotteries that are compared,
it preserves equality of the second dual moments.
Indeed, one may verify that
\begin{align*}
\E{\min(D^{(3)}_1,D^{(3)}_2)}=\E{\min(C^{(3)}_1,C^{(3)}_2)}.
\end{align*}
This is the fundamental reason why EU and DT may diverge from the third order onwards.

We note that if we would have started from an initial lottery $\tilde{L}^{(3)}$ given by
\begin{equation*}\tilde{L}^{(3)}=[\ 1,1/3\ ;\ 2,1/3\ ;\ 3,1/3\ ],\end{equation*}
equality of the variances would have been preserved
in the transformation to the corresponding $\tilde{D}^{(3)}$ and $\tilde{C}^{(3)}$.
In this case both EU DM's with $U'''\geq 0$ and DT DM's with $h'''\geq 0$
would have preferred $\tilde{D}^{(3)}$ over $\tilde{C}^{(3)}$.
That is, EU and DT may, but need not, diverge at the third order.

In order to interpret the sign of the fourth derivative of $h$ we start from an expanded lottery $L^{(4)}$ given by
\begin{equation*}L^{(4)}=[\ 1,1/4\ ;\ 2,1/4\ ;\ 4,1/4\ ;\ 7,1/4\ ].\end{equation*}
Now on the worst three states of $L^{(4)}$ 
we conduct the beneficial transformation
described at the third order (i.e., that with $\mathcal{G}^{(3)}$ preceding $\mathcal{B}^{(3)}$),
attaching
\begin{equation*}\mathcal{G}^{(4)}=[\ x,p\ ;\ -2x,p\ ;\ x,p\ ],\end{equation*}
with in this case $p=1/n$, $n=4$, and $x=1/M$, $M\geq 4$.
Next, on the best three states of $L^{(4)}$
we conduct exactly the opposite transformation,
by attaching
\begin{equation*}\mathcal{B}^{(4)}=[\ -x,p\ ;\ 2x,p\ ;\ -x,p\ ],\end{equation*}
so that we obtain a lottery $D^{(4)}$ given by
\begin{equation*}D^{(4)}=[\ 1+1/M,1/4\ ;\ 2-3/M,1/4\ ;\ 4+3/M,1/4\ ;\ 7-1/M,1/4\ ].\end{equation*}
Conversely, by letting the bad precede the good we generate
\begin{equation*}C^{(4)}=[\ 1-1/M,1/4\ ;\ 2+3/M,1/4\ ;\ 4-3/M,1/4\ ;\ 7+1/M,1/4\ ].\end{equation*}
In Section \ref{Sec:Res}, we show that
$D^{(4)}$ is unanimously preferred over $C^{(4)}$
by DT DM's with ``dual temperance'' (or ``$h''''\leq 0$'')---see Theorems \ref{Th:Char4-1} and \ref{Th:Char4-2}.


To see the implications of the dual story within the EU model,
it is convenient to compare 
$D^{(4)}$
to
$C^{(4)}$
when $M=4$.
While in this case $C^{(4)}$ and $D^{(4)}$ have the same mean and variance,
one easily verifies that $D^{(4)}$ has both a smaller skewness and a smaller kurtosis than $C^{(4)}$.
Hence, there cannot be unanimity among the prudent and temperate EU DM's about the appreciation of the two lotteries:
some will prefer $D^{(4)}$ while others will prefer $C^{(4)}$, depending upon their relative degrees of prudence and temperance.
This observation makes explicit the reason why at the fourth order EU and DT may (continue to) diverge:
while the sequence of squeezing and anti-squeezing at the fourth order may produce different third primal moments for $C^{(4)}$ and $D^{(4)}$,
it preserves the equality of the third dual moments, corroborating again their relevance for the dual story.

We finally note that the lotteries at the third and fourth orders are generated by
simple iterations of the transformations relevant at the second and third orders, respectively.
Hence, this analysis can be pursued up to arbitrary order to interpret the signs of the 
successive derivatives of the probability weighting function.
We thus obtain, as in the Eeckhoudt and Schlesinger (2006) approach, a sequence of simple nested lotteries that yield now the appropriate interpretation
of the signs of $h^{(m)}$.


\setcounter{equation}{0}

\section{Portfolio Choice with Derivatives}\label{Sec:PC}

Consider a DT investor with initial sure wealth $w_{0}$.
Suppose that he allocates an amount $\alpha$ to a risky asset (the ``stock'') and an amount $w_{0}-\alpha$ to a risk-free asset (the ``bond'').
The bond (stock) earns a sure (risky) return of $r$ ($R$) per unit invested.
Assume that $r$ and $R$ are independent of the amounts invested.
The investor aims to determine the optimal amount $\alpha^{*}$.
Because $(w_{0}-\alpha)(1+r)+\alpha(1+R)=w_{0}(1+r)+\alpha(R-r)$
and the DT evaluation is translation invariant and positively homogeneous,
his problem reads:
\begin{equation}
\label{eq:PC}
\argmax_{\alpha}\{\alpha(V\left[R\right]-r)\}.
\end{equation}
We add the constraint $0\leq\alpha\leq w_{0}$.
It readily follows that the optimal solution $\alpha^{*}$ is a corner solution.

Suppose $V\left[R\right]<r$.
Then the optimal solution is $\alpha^{*}=0$.
In this case, the investor fully invests in the bond.
Now imagine that the investor is offered an improvement to the distribution of the risky return.
In particular, he is offered the possibility of supplementing the stock with
derivative products on the stock that have zero expected value at zero cost.
The derivative products
are selected by applying the dual story
such that they induce, through squeezing and anti-squeezing,
an $m^{\mathrm{th}}$ order improvement
of $R$, with $m\geq 2$.
The return of the risky portfolio of stock and derivatives jointly is denoted by $\bar{R}$.
(We provide details on the derivative products shortly.)

Then two cases are possible:
(i) If $V[\bar{R}]\leq r$, full investment in the bond remains optimal.
(ii) If the improvement is sufficiently large so that $V[\bar{R}]>r$, the DT investor will shift from one corner solution
(full investment in the bond, i.e., $\alpha^{*}=0$)
to the other corner solution
(full investment in the risky portfolio, i.e., $\alpha^{*}=w_{0}$).

\begin{landscape}
\begin{figure}[H]
\begin{center}
\vskip -1.58cm
\caption{Optimal Portfolio Choice with Derivatives.
\newline\footnotesize This figure plots the payoff functions of the derivative products
to supplement the stock
inducing an $m^{\mathrm{th}}$ order improvement, $m=2,3,4$.
At each order, the derivative products have zero (joint) expected value.
At the second order (top left panel), we supplement the stock with a long position in a put option (blue, solid)
and a short position in a call option (green, dashed), both with a strike price of 2.
At the third order (top right panel), we enter into a straddle option (blue)
at stock price 4.
Finally, at the fourth order (bottom panel), we add a long straddle at stock price 4 (blue)
and a short straddle at stock price 12 (green, dashed).
Both the stock price itself and the stock-and-derivatives portfolio price can at each order be generated
from a common initial lottery (in the form of a stock-and-derivatives portfolio)
represented by the grey dotted lines 
through (a sequence of) squeezes and anti-squeezes represented by the solid and dashed arrows, respectively, according to the dual story;
see also footnotes \ref{fn:pc1} and \ref{fn:pc2}.
}
\vskip 0.4cm
\includegraphics[scale=0.56,angle=0]{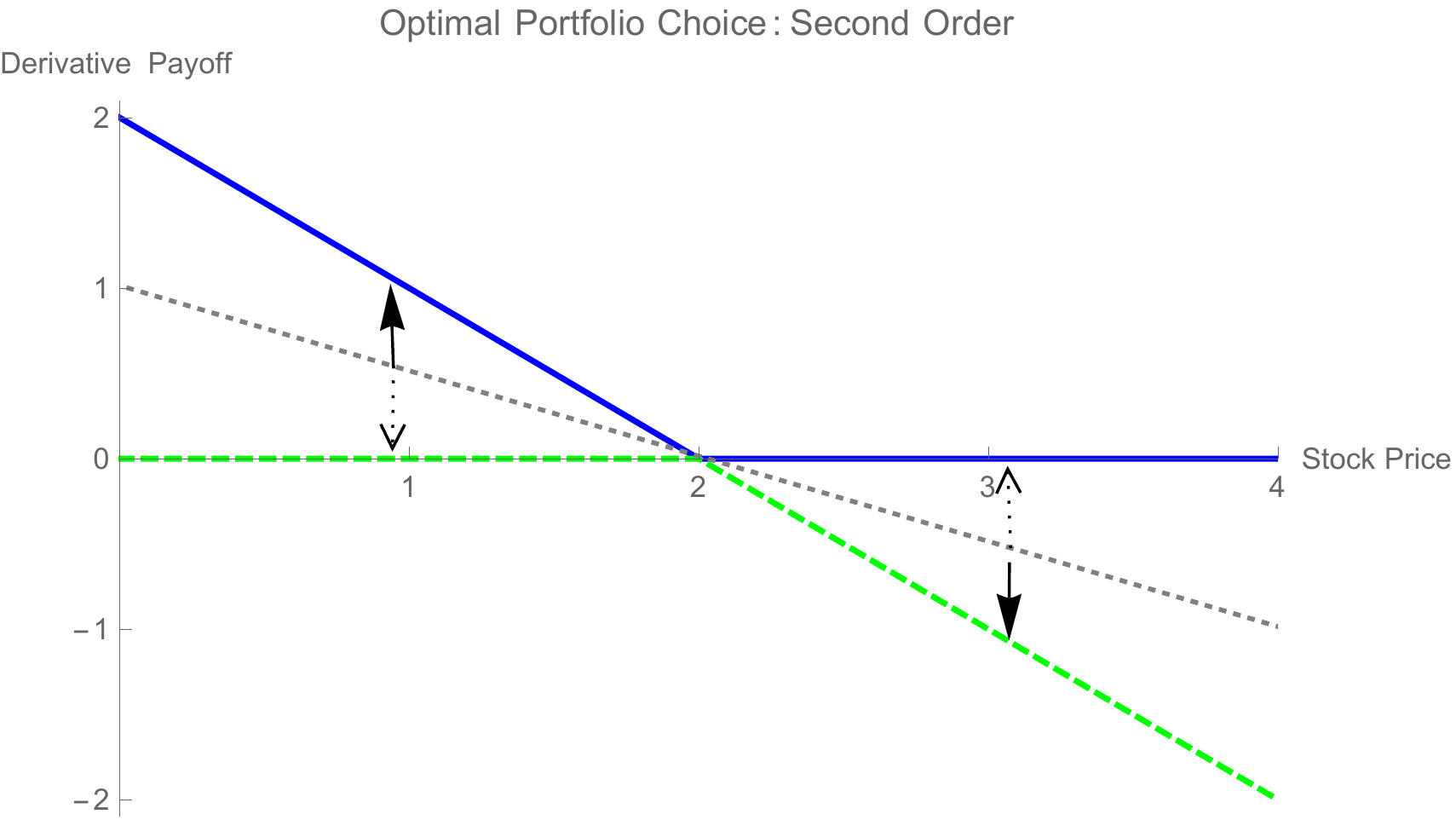}
\includegraphics[scale=0.56,angle=0]{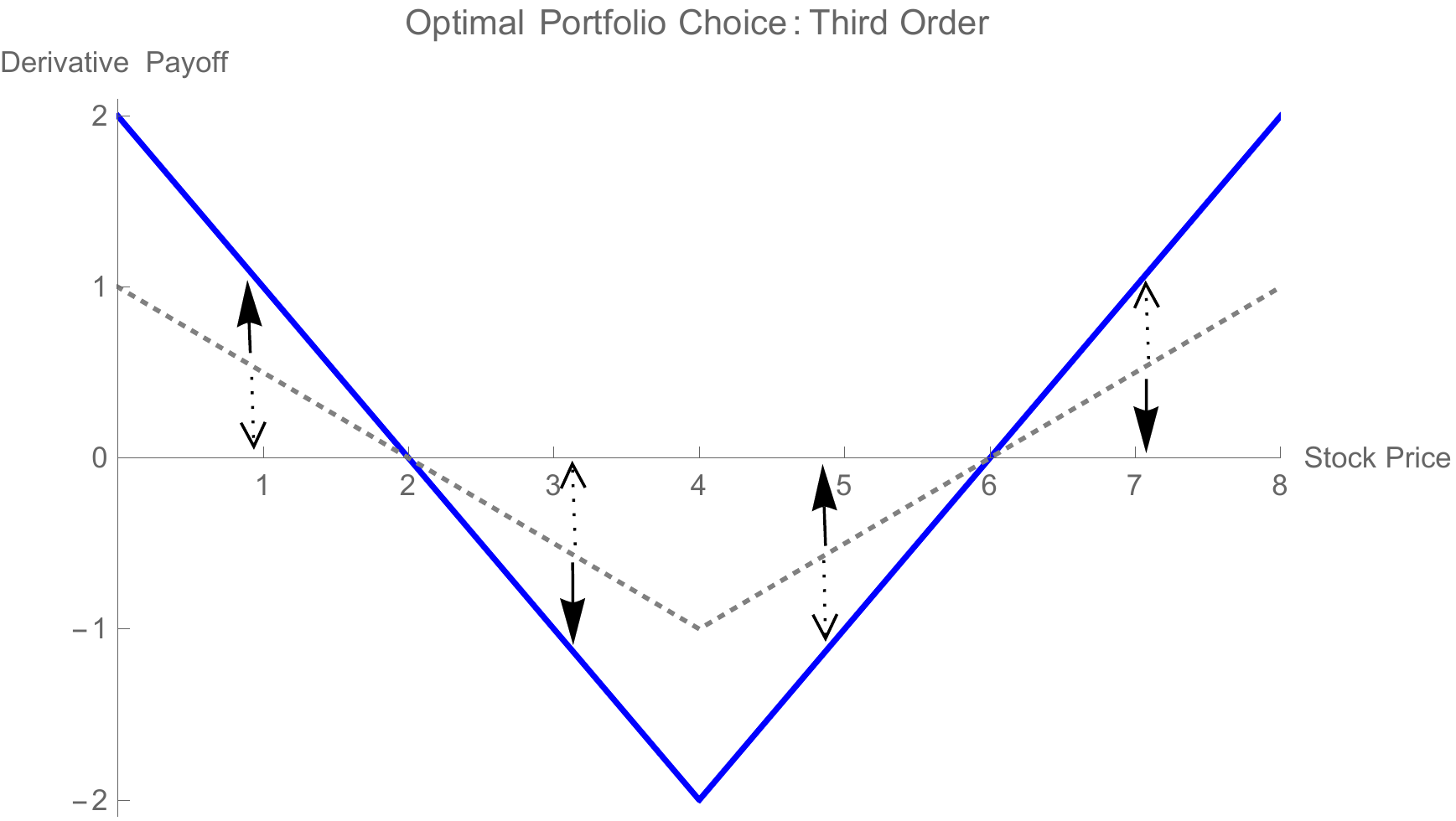}\vskip 0.3cm
\includegraphics[scale=0.56,angle=0]{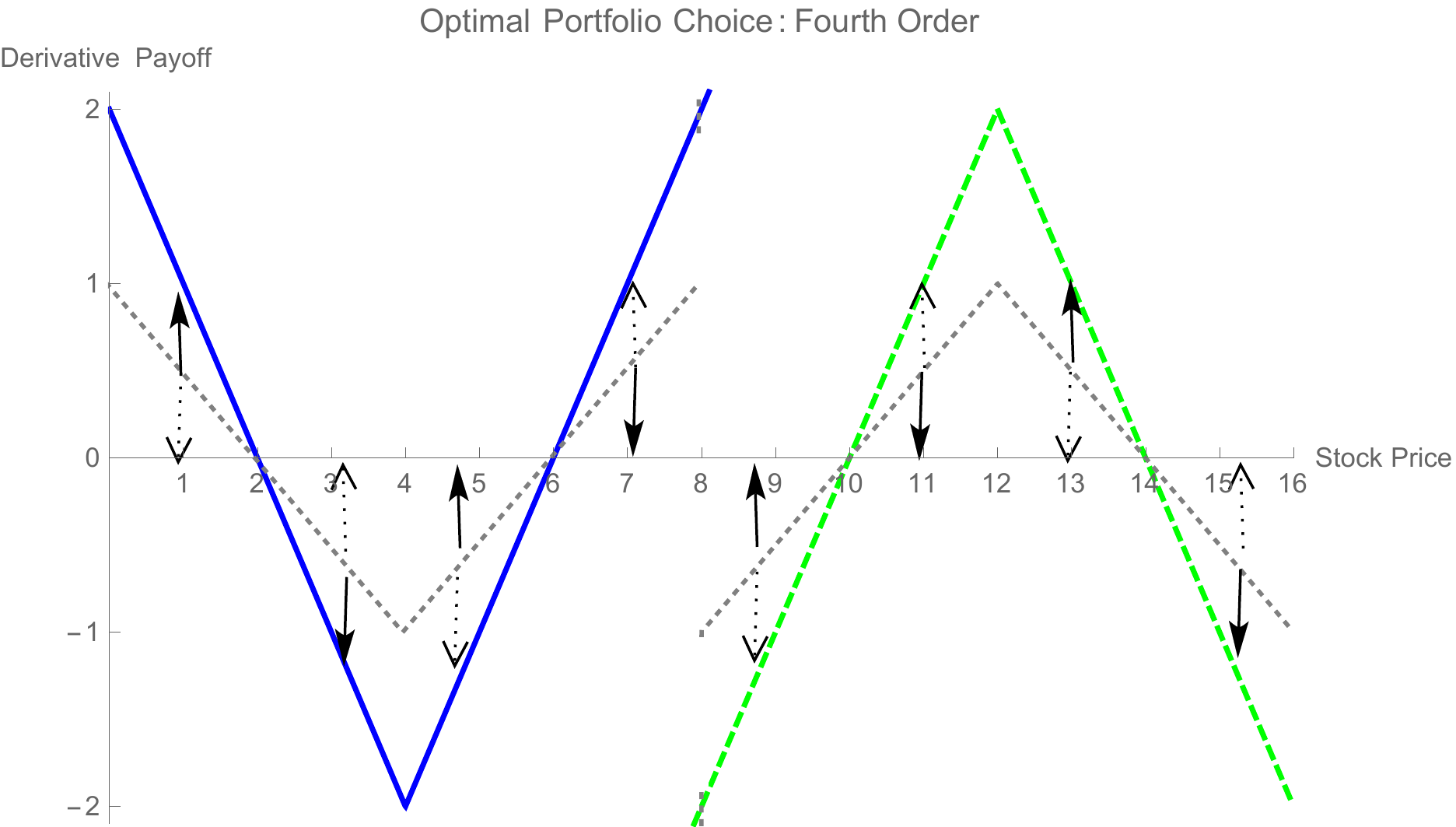}
\label{fig:PCder}
\end{center}
\end{figure}
\end{landscape}

To illustrate these $m^{\mathrm{th}}$ order
improvements, 
suppose for ease of exposition that $r\equiv 0$.
Consider the second order first.
Assume that 
the stock price takes the values 1 and 3 each with probability 1/2.
Invoking the dual story,
we find that
a long position in a put option 
combined with a short position in a call option, 
each with a strike price of 2 such that the joint expected value is zero,
improves the attractiveness of the risky portfolio ($\bar{R}$ versus $R$)
whenever 
$h''\leq 0$.
Indeed, the difference between $\bar{R}$ and $R$ can in this case be seen as the result of squeezing and anti-squeezing.\footnote{\label{fn:pc1}Throughout this section
the risky stock price $S_{0}(1+R)$, with $S_{0}$ the initial stock price,
plays the role of the lotteries $C^{(m)}$ encountered in the dual story,
while the stock-and-derivatives portfolio price $S_{0}(1+\bar{R})$ plays the role of the lotteries $D^{(m)}$.
The
derivative supplements directly generate the improvement $\bar{R}$ from $R$.
Specifically,
$R=(C^{(m)}-S_{0})/S_{0}$ and $\bar{R}=(D^{(m)}-S_{0})/S_{0}$.
At the second order,
\begin{equation*}C^{(2)}=[\ 1,1/2\ ;\ 3,1/2\ ]\qquad\mathrm{and}\qquad D^{(2)}=[\ 2,1/2\ ;\ 2,1/2\ ].\end{equation*}
Of course, $C^{(2)}$ and $D^{(2)}$ may be generated from a common initial lottery $L^{(2)}=[\ 3/2,1/2\ ;\ 5/2,1/2\ ]$
by squeezing and anti-squeezing, that is, by attaching 
\begin{equation*}\mathcal{G}^{(2)}=[\ 1/2,1/2\ ]\qquad\mathrm{and}\qquad \mathcal{B}^{(2)}=[\ -1/2,1/2\ ],\end{equation*}
to $L^{(2)}$.
If the good (bad) precedes the bad (good), $D^{(2)}$ ($C^{(2)}$) is generated.}
As visualized in the top left panel of Figure \ref{fig:PCder},
this combination of long put and short call provides a hedge against adverse stock scenarios,
which is financed by giving up upward potential.

Next, consider the third order.
Assume now that 
the stock price takes the values $\{1, 3, 5, 7\}$ each with probability 1/4.
Then, deploying the dual story,
one may readily verify that a so-called ``straddle'' option at stock price 4
such that the expected value is zero,
improves the attractiveness of the risky portfolio ($\bar{R}$ versus $R$) whenever 
$h'''\geq 0$.
As visualized in the top right panel of Figure \ref{fig:PCder},
the straddle pays off in bad stock scenarios and in good stock scenarios,
but generates losses in intermediate stock scenarios.\footnote{\label{fn:pc2}At the third order,
$R=(C^{(3)}-S_{0})/S_{0}$ and $\bar{R}=(D^{(3)}-S_{0})/S_{0}$,
where
\begin{equation*}C^{(3)}=[\ 1,1/4\ ;\ 3,1/4\ ;\ 5,1/4\ ;\ 7,1/4\ ]\qquad\mathrm{and}\qquad
D^{(3)}=[\ 2,1/4\ ;\ 2,1/4\ ;\ 4,1/4\ ;\ 8,1/4\ ].\end{equation*}
They can be generated from an initial lottery $L^{(3)}=[\ 3/2,1/4\ ;\ 5/2,1/4\ ;\ 9/2,1/4\ ;\ 15/2,1/4\ ]$ by attaching
\begin{equation*}\mathcal{G}^{(3)}=[\ 1/2,1/4\ ;\ -1/2,1/4\ ]\qquad\mathrm{and}\qquad \mathcal{B}^{(3)}=[\ -1/2,1/4\ ;\ 1/2,1/4\ ],\end{equation*}
at non-overlapping states.
}
Because the straddle increases not only the skewness but also the variance of the risky portfolio,
there would be no unanimity for the straddle supplement under EU, contrary to under DT.

Finally, at the fourth order,
assume that the stock price takes the values $\{1, 3, 5, 7,$ $9, 11, 13, 15\}$ each with probability 1/8.
Then, the dual story implies that a long straddle at stock price 4
combined with a short straddle at stock price 12,
with joint expected value equal zero,
improves the attractiveness of the risky portfolio 
provided 
$h''''\leq 0$.
This combination of a long and short straddle, visualized in the last panel of Figure \ref{fig:PCder},
is a popular and simple case of a so-called ``volatility spread''.\footnote{The payoffs of the long and short straddle are digitally set to 0 at stock price 8 for simplicity.}

We thus find that an $m^{\mathrm{th}}$ order (dual) improvement of the risky asset's return distribution
never reduces the demand for the risky asset
when the successive derivatives of the probability weighting function alternate in sign.
This stands in sharp contrast to the familiar results under EU,
where an $m^{\mathrm{th}}$ order (primal) improvement of $R$
has ambiguous demand effects, even for $m\geq 1$.
Indeed, to obtain the natural result that such improvements induce a higher demand for the risky asset under EU,
additional non-trivial conditions need to be imposed
(see Section 9.3 of Eeckhoudt and Gollier (1995) and Gollier (1995) for details).\footnote{See also
the analysis of Dittmar (2002) in the EU model.}

\setcounter{equation}{0}

\section{Self-Protection with Background Risk}\label{Sec:Prot}

In this section, we consider a DT DM with initial sure wealth $w_{0}$ who faces the risk of losing the monetary amount $\ell >0$.
The probability of occurrence of the loss, $p$, depends on the self-protection\footnote{Adopting the terminology of Ehrlich and Becker (1972).}
effort,
$e$, exerted by the DM.
In particular, $p(e)$ is decreasing in $e$. 
Effort is measured in monetary equivalents.
The DM aims to determine the optimal level of effort $e^{*}$ that maximizes his DT evaluation.

We analyze this problem in the presence of an independent background risk.
This background risk may e.g., be due to the risk of holding risky financial assets or to uncertain labor income.
The sign of $h'''$ appears to play an essential role in this self-protection problem with background risk.
Hence, while under EU the a priori validity of $U'''\geq 0$ is connected to a savings decision (Kimball (1990)),
we show that under DT the a priori validity of $h'''\geq 0$ may be linked to a self-protection problem.

With an independent binary background risk taking the value $\pm\varepsilon$, $\varepsilon>0$, each with probability $1/2$,
the self-protection problem can be represented by the following 4-state lottery $S(e)$:
\begin{align*}S(e)=[\ &w_{0}-\ell -\varepsilon-e,p(e)/2\ ;\ w_{0}-\ell +\varepsilon-e,p(e)/2\ ;\ \\&w_{0}-\varepsilon-e,(1-p(e))/2\ ;\ w_{0}+\varepsilon-e,(1-p(e))/2\ ],
\end{align*}
assuming $2\varepsilon<\ell $.
If $2\varepsilon>\ell $, the middle two states of $S$ change places.
The DT DM solves:
\begin{equation}
\label{eq:Prot}
\argmax_{e}\{V\left[S(e)\right]\}.
\end{equation}
We assume throughout that $V$ is concave in $e$.

If $2\varepsilon<\ell $,
the first-order condition for optimality reads
\begin{align*}
\frac{V\left[S(e)\right]}{\mathrm{d}e}=&\ (p'(e)/2)h'(p(e)/2)\left((w_{0}-\ell -\varepsilon-e)-(w_{0}-\ell +\varepsilon-e)\right)\\
&\ +p'(e)h'(p(e))\left((w_{0}-\ell +\varepsilon-e)-(w_{0}-\varepsilon-e)\right)\\
&\ +(p'(e)/2)h'((1+p(e))/2)\left((w_{0}-\varepsilon-e)-(w_{0}+\varepsilon-e)\right)\\
&\ -1=0.
\end{align*}
After obvious simplifications, we obtain
\begin{align}\label{eq:ProtFOC1}p'(e)\varepsilon 
\left(-h'(p(e)/2)+2h'(p(e))-h'((1+p(e))/2)\right)
-p'(e)h'(p(e))\ell -1=0.
\end{align}
We note that in the absence of any background risk, i.e., when $\varepsilon =0$, the first-order condition reduces to
\begin{equation}\label{eq:ProtFOCNoBr}-p'(e)h'(p(e))\ell -1=0.\end{equation}
Now assume first, like in Eeckhoudt and Gollier (2005) for EU, that
$-p'(e)h'(p(e))\ell -1=0$ when $p(e)=1/2$.
This means that the DT DM optimally selects an effort level $e$
such that the probability of the occurrence of a loss is $1/2$,
in the absence of background risk.
Under this assumption, we find from (\ref{eq:ProtFOC1}) that the impact of the background risk is linked to the sign of
\begin{equation}\label{eq:h3rdhalf}
-h'(1/4)+2h'(1/2)-h'(3/4)=\left(h'(1/2)-h'(1/4)\right)-\left(h'(3/4)-h'(1/2)\right).\end{equation}
If the sign of $h'''$ is positive, then \eqref{eq:h3rdhalf} is negative, and note that $p'(e)\varepsilon$ is negative, too.
Therefore, and in view of the concavity of $V$ with respect to $e$,
the introduction of a background risk in this setting stimulates self-protection provided the DT DM is dual prudent. 

Next, consider the case $2\varepsilon>\ell $.
In this case,
the first-order condition is
\begin{align*}
\frac{\mathrm{d}V\left[S(e)\right]}{\mathrm{d}e}=&\ (p'(e)/2)h'(p(e)/2)\left((w_{0}-\ell -\varepsilon-e)-(w_{0}-\varepsilon-e)\right)\\
&\ +(p'(e)/2)h'((1+p(e))/2)\left((w_{0}-\ell +\varepsilon-e)-(w_{0}+\varepsilon-e)\right)\\
&\ -1=0,
\end{align*}
which, after obvious simplifications, reduces to
\begin{equation}
\label{eq:ProtFOC2}
-(1/2)p'(e)\ell  \left(h'(p(e)/2)+h'((1+p(e))/2)\right)-1=0.
\end{equation}
Recall that in the absence of any background risk, the first-order condition equals (\ref{eq:ProtFOCNoBr}).
Hence, noting that the difference between the left-hand side of \eqref{eq:ProtFOC2}
and the left-hand side of \eqref{eq:ProtFOCNoBr} equals
\begin{align}\label{eq:ProtFOC2-FOCNoBr}
-(1/2)p'(e)\ell \left(h'(p(e)/2)-2h'(p(e))+h'((1+p(e))/2)\right),
\end{align}
and maintaining the assumption that $-p'(e)h'(p(e))\ell -1=0$ when $p(e)=1/2$,
we find that the impact of the background risk is again linked to the sign of (\ref{eq:h3rdhalf})
which, in turn, depends on the sign of $h'''$.

Thus, under the maintained assumption that $-p'(e)h'(p(e))\ell -1=0$ when $p(e)=1/2$,
$h'''\geq 0$ guarantees that the marginal benefit of self-protection increases
upon the introduction of an independent background risk.
Therefore, the background risk stimulates self-protection under dual prudence.\footnote{There is a difference between the two cases.
When $2\varepsilon<\ell $, the convexity of $h'$ adds a marginal benefit that depends on the size of $\varepsilon$
compared to the situation of no background risk; cf. (\ref{eq:ProtFOC1}).
When $2\varepsilon>\ell $, the convexity of $h'$ adds a marginal benefit that is independent of $\varepsilon$ but depends on the size of $\ell $
compared to the situation of no background risk; cf. (\ref{eq:ProtFOC2-FOCNoBr}).}

By a similar analysis, we can extend the results to a larger set of possible cases.\footnote{See the extended online version for further details.}
If $-p'(e)h'(p(e))\ell -1=0$ when $p(e)\leq 1/2$, we find that
dual imprudent DM's (i.e., $h'''\leq 0$) exert less effort with background risk than without.
If $-p'(e)h'(p(e))\ell -1=0$ when $p(e)\geq 1/2$, we obtain that
dual prudent DM's (i.e., $h'''\geq 0$) exert more effort with background risk than without.

%

\setcounter{equation}{0}

\section{Dual Risk Apportionment}\label{Sec:Res}

In this section, we develop the dual story as illustrated in Section \ref{Sec:Pref} in full generality.
Throughout this section we consider $n$-state lotteries
and assume all states to have equal probability of occurrence $1/n$, $n\in\mathbb{N}$.
We permit 
that the increments in the outcomes when moving to an adjacent state are equal to zero.
This can be interpreted to yield lotteries with unequal state probabilities.
%
%

\subsection{The Third Order: Dual Prudence}\label{sec:Res3rd}

We start by providing a simple class of lottery pairs
such that the direction of preference between these lottery pairs is equivalent
to signing the third derivative of the probability weighting function under DT.
Consider
\begin{equation*}\mathcal{G}^{(3)}=[\ \delta,1/n\ ;\ \ldots ;\ -\delta,1/n\ ]\quad\mathrm{and}\quad
\mathcal{B}^{(3)}=[\ -\delta',1/n\ ;\ \ldots ;\ \delta',1/n\ ],\end{equation*}
with $\delta,\delta'\geq 0$.
The acronyms ``$\mathcal{G}$'' and ``$\mathcal{B}$'' refer to ``good'' and ``bad''.
Upon ``attaching'' $\mathcal{G}^{(3)}$ and $\mathcal{B}^{(3)}$ state-wise
to an arbitrarily given initial lottery
that has at least three states such that the state-wise addition is feasible,
we generate $D^{(3)}$ if $\mathcal{G}^{(3)}$ state-wise precedes $\mathcal{B}^{(3)}$
and we generate $C^{(3)}$ if $\mathcal{B}^{(3)}$ state-wise precedes $\mathcal{G}^{(3)}$.
Attaching the good, $\mathcal{G}^{(3)}$, induces a squeeze,
while attaching the bad, $\mathcal{B}^{(3)}$, induces an anti-squeeze.

To emphasize the generality of these transformations,
we note that: (i) $\delta$ and $\delta'$ may differ;
(ii) the number of states that ``$\ldots$'' denotes in $\mathcal{G}^{(3)}$ and in $\mathcal{B}^{(3)}$ may differ;
(iii) the number of states that ``$\ldots$'' denotes in $\mathcal{G}^{(3)}$ and in $\mathcal{B}^{(3)}$ 
may be equal to zero;
(iv) the spacing (i.e., the number of states) between $\mathcal{G}^{(3)}$ and $\mathcal{B}^{(3)}$ on the one hand
and $\mathcal{B}^{(3)}$ and $\mathcal{G}^{(3)}$ on the other hand
when attaching them to the initial lottery
may differ;
(v) this spacing may be negative, so that the good and the bad partially overlap,
but $\mathcal{G}^{(3)}$ strictly precedes $\mathcal{B}^{(3)}$ by at least one state for $D^{(3)}$
and $\mathcal{B}^{(3)}$ strictly precedes $\mathcal{G}^{(3)}$ by at least one state for $C^{(3)}$.
All these generalities are permitted as long as the 
ranking of outcomes remains unaffected
by the squeezes and anti-squeezes
and the outcomes of the resulting lotteries remain non-negative.
These generalities are in part thanks to the fact that we don't compare the changed lotteries to the initial lottery they are generated from.
Rather we compare two altered lotteries: one has the good precede the bad, the other the reverse.

We say that an individual is ``dual prudent'' if, for any such lottery pair $(C^{(3)},D^{(3)})$, he prefers $D^{(3)}$ to $C^{(3)}$.
The following theorem shows that within DT this is guaranteed by a positive sign of $h'''$.
The individual's type of behavior corresponding to these lottery preferences,
and their higher-order generalizations that we develop below,
may be termed ``dual risk apportionment''.
Much like under primal risk apportionment,
the apportionment of harms within a lottery mitigates the detrimental effects
for such an individual.

\begin{theorem}\label{Th:Char3-1}
If $C^{(3)}$ and $D^{(3)}$ are generated by the transformations described above,
then $D^{(3)}$ is preferred (dispreferred) to $C^{(3)}$ by any DT DM with $h'''\geq 0$ ($h'''\leq 0$).
\end{theorem}

In the remainder of this subsection,
we consider a parsimonious subclass of lottery pairs $C^{(3)}$ and $D^{(3)}$
that turns out to be already sufficient
for the purpose of
signing $h'''$.
Let $n\geq 3$ and consider
\begin{align*}C^{(3)}_{n}=&\ [\ x_{1},1/n\ ;\ \ldots ;\ x_{j},1/n\ ;\\
&\quad\ \mathbf{x_{j+1},1/n}\ ;\ \mathbf{x_{j+2},1/n}\ ;\ \mathbf{x_{j+3},1/n}\ ;\\
&\quad\qquad\qquad\qquad\qquad\qquad\ x_{j+4},1/n; \ \ldots\ ;\ .,1/n\ ].\end{align*}
The three states in bold need to be present for any given $n (\geq 3)$,
the remaining states are added arbitrarily until the state probabilities sum up to 1.
The increments in the outcomes when moving to an adjacent higher state of $C^{(3)}_{n}$ are allowed to be arbitrarily non-negative and state-dependent,
as long as 
the squeezes and anti-squeezes performed below do not change the ranking of outcomes.\footnote{This is accomplished by restricting $M$ in the squeezes and anti-squeezes explicated below accordingly.}

Then, we (directly) construct $D^{(3)}_{n}$ from $C^{(3)}_{n}$ by attaching to the three states in bold
\begin{equation*}C2D^{(3)}_{n}=[\ 1/M,1/n\ ;\ -2/M,1/n\ ;\ 1/M,1/n\ ].\end{equation*}
This yields $D^{(3)}_{n}$ given by
\begin{align*}D^{(3)}_{n}=&\ [\ x_{1},1/n\ ;\ \ldots ;\ x_{j},1/n\ ;\\
&\quad\ \mathbf{x_{j+1}+1/M,1/n}\ ;\ \mathbf{x_{j+2}-2/M,1/n}\ ;\ \mathbf{x_{j+3}+1/M,1/n}\ ;\\
&\qquad\qquad\qquad\qquad\qquad\qquad\qquad\ x_{j+4},1/n; \ \ldots\ ;\ .,1/n\ ].
\end{align*}

Clearly, $C^{(3)}_{n}$ and $D^{(3)}_{n}$ occur as a subclass of $C^{(3)}$ and $D^{(3)}$.
To see this, consider an initial lottery
$L^{(3)}_{n}$ given by
\begin{align*}L^{(3)}_{n}=&\ [\ x_{1},1/n\ ;\ \ldots ;\ x_{j},1/n\ ;\\
&\quad\ \mathbf{x_{j+1}+1/(2M),1/n}\ ;\ \mathbf{x_{j+2}-1/M,1/n}\ ;\ \mathbf{x_{j+3}+1/(2M),1/n}\ ;\\
&\quad\qquad\qquad\qquad\qquad\qquad\ x_{j+4},1/n; \ \ldots\ ;\ .,1/n\ ],\end{align*}
and generate $D^{(3)}_{n}$ ($C^{(3)}_{n}$) by attaching the good (bad), $\mathcal{G}^{(3)}$ ($\mathcal{B}^{(3)}$), given by
\begin{equation*}\mathcal{G}^{(3)}=[\ 1/(2M),1/n\ ;\ -1/(2M),1/n\ ]\quad\mathrm{and}\quad \mathcal{B}^{(3)}=[\ -1/(2M),1/n\ ;\ 1/(2M),1/n\ ],\end{equation*}
to the first two bold states of $L^{(3)}_{n}$,
and attaching the bad (good), $\mathcal{B}^{(3)}$ ($\mathcal{G}^{(3)}$),
to the last two bold states of $L^{(3)}_{n}$.
To keep the required number of states as small as possible and enhance parsimony,
sufficient for our purpose of signing $h'''$,
we squeeze and anti-squeeze at an overlapping state, just like in Section \ref{Sec:Pref}.


The following theorem shows that within DT the preference towards the class of lottery pairs $D^{(3)}_{n}$ and $C^{(3)}_{n}$
signs $h'''$:
\begin{theorem}\label{Th:Char3-2}
If, for any $n\geq 3$, a DT DM prefers (disprefers) $D^{(3)}_{n}$ to $C^{(3)}_{n}$,
then $h'''\geq 0$ ($h'''\leq 0$).
\end{theorem}

\subsection{The Fourth Order: Dual Temperance}\label{sec:Res4th}

Next, we provide a simple class of lottery pairs
such that the direction of preference between these lottery pairs is equivalent
to signing the fourth derivative of the probability weighting function under DT.
Consider
\begin{align*}
\mathcal{G}^{(4)}&=[\ \delta,1/n\ ;\ \ldots ;\ -\delta,1/n\ ;\ \ldots ;\ -\delta,1/n\ ;\ \ldots ;\ \delta,1/n\ ]\quad \mathrm{and}\nonumber\\
\mathcal{B}^{(4)}&=[\ -\delta',1/n\ ;\ \ldots ;\ \delta',1/n\ ;\ \ldots ;\ \delta',1/n\ ;\ \ldots ;\ -\delta',1/n\ ],\end{align*}
with $\delta,\delta'\geq 0$.
Upon attaching $\mathcal{G}^{(4)}$ and $\mathcal{B}^{(4)}$ to an arbitrarily given initial lottery
that has at least four states such that the state-wise addition is feasible,
we generate $D^{(4)}$ if $\mathcal{G}^{(4)}$ state-wise precedes $\mathcal{B}^{(4)}$
and we generate $C^{(4)}$ if $\mathcal{B}^{(4)}$ state-wise precedes $\mathcal{G}^{(4)}$.
We note that $\mathcal{G}^{(4)}$ and $\mathcal{B}^{(4)}$ can each be interpreted as simple concatenations of
same-sized but exactly opposite $\mathcal{G}^{(3)}$ and $\mathcal{B}^{(3)}$.
Thus, we define the fourth order transformations by simple iterations
of the third order ones.

To emphasize the generality of these transformations, we note that:
(i) $\delta$ and $\delta'$ may differ;
(ii) the number of states that ``$\ldots$'' denotes may differ within and between $\mathcal{G}^{(4)}$ and $\mathcal{B}^{(4)}$,
but both $\mathcal{G}^{(4)}$ and $\mathcal{B}^{(4)}$ need to be symmetrical
in the sense that the first and third ``$\ldots$'' within $\mathcal{G}^{(4)}$ denote the same number of states
and similarly for $\mathcal{B}^{(4)}$
(which is compatible with $\mathcal{G}^{(4)}$ and $\mathcal{B}^{(4)}$ each being
concatenations of same-sized but exactly opposite $\mathcal{G}^{(3)}$ and $\mathcal{B}^{(3)}$);
(iii) the number of states that the first and third ``$\ldots$'' denote in $\mathcal{G}^{(4)}$ and $\mathcal{B}^{(4)}$ is non-negative and
may be equal to zero,
while the number of states that the middle (second) ``$\ldots$'' denotes is larger than or equal to minus one;\footnote{If the number of states that
``$\ldots$'' represents is -1, the middle two adjacent states overlap, thus attaching $-2\delta$ for $\mathcal{G}^{(4)}$ and $2\delta'$ for $\mathcal{B}^{(4)}$ to a state with probability $1/n$.}
(iv) the spacing (i.e., the number of states) between $\mathcal{G}^{(4)}$ and $\mathcal{B}^{(4)}$ on the one hand
and $\mathcal{B}^{(4)}$ and $\mathcal{G}^{(4)}$ on the other hand when attaching them to the initial lottery may differ;
(v) this spacing may be negative, so that the good and the bad partially overlap,
but $\mathcal{G}^{(4)}$ strictly precedes $\mathcal{B}^{(4)}$ by at least one state for $D^{(4)}$
and $\mathcal{B}^{(4)}$ strictly precedes $\mathcal{G}^{(4)}$ by at least one state for $C^{(4)}$.
All these generalities are permitted as long as the ranking of outcomes remains unaffected
and the outcomes of the resulting lotteries remain non-negative.

We say that an individual is ``dual temperate'' if, for any such lottery pair $(C^{(4)},D^{(4)})$,
he prefers $D^{(4)}$ to $C^{(4)}$.
The following theorem shows that within DT this is guaranteed
by a negative sign of $h''''$.

\begin{theorem}
\label{Th:Char4-1}
If $C^{(4)}$ and $D^{(4)}$ are generated by the transformations described above,
then
$D^{(4)}$ is preferred (dispreferred) to $C^{(4)}$ by any DT DM with $h''''\leq 0$ ($h''''\geq 0$).
\end{theorem}

As in Section \ref{sec:Res3rd}, let us in the remainder of this subsection consider a parsimonious subclass of lottery pairs
$C^{(4)}$ and $D^{(4)}$,
already sufficient for our purpose of signing $h''''$.
Let $n\geq 4$ and consider
\begin{align*}C^{(4)}_{n}=\ &[\ x_{1},1/n\ ;\ \ldots ;\ x_{j},1/n\ ;\\
&\quad\ \mathbf{x_{j+1},1/n}\ ;\ \mathbf{x_{j+2},1/n}\ ;\ \mathbf{x_{j+3},1/n}\ ;\ \mathbf{x_{j+4},1/n}\ ;\\
&\quad\qquad\qquad\qquad\qquad\qquad\ x_{j+5},1/n; \ \ldots\ ;\ .,1/n\ ].\end{align*}
The four states in bold need to be present for any given $n (\geq 4)$, the remaining states are added arbitrarily until the state probabilities sum up to 1.
The increments in the outcomes when moving to an adjacent higher state of $C^{(4)}_{n}$ are again allowed to be arbitrarily non-negative and state-dependent,
as long as 
the transformations conducted below do not change the ranking of outcomes.\footnote{This is accomplished by restricting $M$ in the transformations explicated below accordingly.}

Then, we (directly) construct $D^{(4)}_{n}$ from $C^{(4)}_{n}$ by attaching to the four states in bold
\begin{equation*}C2D^{(4)}_{n}=[\ 1/M,1/n\ ;\ -3/M,1/n\ ;\ 3/M,1/n\ ;\ -1/M,1/n\ ].\end{equation*}
This yields $D^{(4)}_{n}$ given by
\begin{align*}D^{(4)}_{n}=\ &[\ x_{1},1/n\ ;\ \ldots ;\ x_{j},1/n\ ;\\
&\quad\ \mathbf{x_{j+1}+1/M,1/n}\ ;\ \mathbf{x_{j+2}-3/M,1/n}\ ;\ \mathbf{x_{j+3}+3/M,1/n}\ ;\ \mathbf{x_{j+4}-1/M,1/n}\ ;\\
&\qquad\qquad\qquad\qquad\qquad\qquad\qquad\ x_{j+5},1/n; \ \ldots\ ;\ .,1/n\ ].
\end{align*}

Clearly, $C^{(4)}_{n}$ and $D^{(4)}_{n}$ occur as a subclass of $C^{(4)}$ and $D^{(4)}$.
To see this, consider an initial lottery $L^{(4)}_{n}$ given by
\begin{align*}L^{(4)}_{n}=\ &[\ x_{1},1/n\ ;\ \ldots ;\ x_{j},1/n\ ;\\
&\ \mathbf{x_{j+1}+1/(2M),1/n}\ ;\ \mathbf{x_{j+2}-3/(2M),1/n}\ ;\ \mathbf{x_{j+3}+3/(2M),1/n}\ ;\\
&\qquad\qquad\qquad\qquad\qquad\qquad\qquad\ \mathbf{x_{j+4}-1/(2M),1/n}\ ;\ x_{j+5},1/n; \ \ldots\ ;\ .,1/n\ ],
\end{align*}
and generate $D^{(4)}_{n}$ ($C^{(4)}_{n}$) by attaching the good (bad), $\mathcal{G}^{(4)}$ ($\mathcal{B}^{(4)}$), given by
\begin{align*}
\mathcal{G}^{(4)}&=[\ 1/(2M),1/n\ ;\ -1/M,1/n\ ;\ 1/(2M),1/n\ ]\quad \mathrm{and}\\
\mathcal{B}^{(4)}&=[\ -1/(2M),1/n\ ;\ 1/M,1/n\ ;\ -1/(2M),1/n\ ],
\end{align*}
to the first three bold states of $L^{(4)}_{n}$,
and attaching the bad (good), $\mathcal{B}^{(4)}$ ($\mathcal{G}^{(4)}$),
to the last three bold states of $L^{(4)}_{n}$.
To keep the required number of states as small as possible and enhance parsimony,
sufficient for our purpose of signing $h''''$,
we again conduct transformations at overlapping states,
just like in Section \ref{sec:Res3rd} and Section \ref{Sec:Pref}.

The following theorem shows that within DT the preference towards the class of lottery pairs $D^{(4)}_{n}$ and $C^{(4)}_{n}$
signs $h''''$:
\begin{theorem}\label{Th:Char4-2}
If, for any $n\geq 4$, a DT DM prefers (disprefers) $D^{(4)}_{n}$ to $C^{(4)}_{n}$,
then $h''''\leq 0$ ($h''''\geq 0$).
\end{theorem}

\subsection{The $m^{\mathrm{th}}$ Order}\label{sec:Resmth}

In this section, we systematically construct simple nested classes of lottery pairs
such that the direction of preference between these lottery pairs
is equivalent to signing the $m^{\mathrm{th}}$ derivative of the probability weighting function under DT.
We start at the second order
and proceed to construct all higher orders of dual risk apportionment by simple iteration, as follows.
Consider
\begin{equation*}
\mathcal{G}^{(2)}=[\ \delta,1/n\ ]\quad\mathrm{and}\quad \mathcal{B}^{(2)}=[\ -\delta',1/n\ ],
\end{equation*}
with $\delta,\delta'\geq 0$.
Taking the $(m-1)^{\mathrm{th}}$ order as a starting point,
we construct the class of lottery pairs for the $m^{\mathrm{th}}$ order in two steps.
Let $n\geq m$.
Then:
\begin{itemize}
\item[(i)] Concatenate a same-sized but exactly opposite $\mathcal{G}^{(m-1)}$ and $\mathcal{B}^{(m-1)}$,
with $\mathcal{G}^{(m-1)}$ state-wise preceding $\mathcal{B}^{(m-1)}$ by at least one state.
This generates $\mathcal{G}^{(m)}$.
Similarly, concatenate another same-sized but exactly opposite $\mathcal{G}^{(m-1)}$ and $\mathcal{B}^{(m-1)}$,
with $\mathcal{B}^{(m-1)}$ state-wise preceding $\mathcal{G}^{(m-1)}$ by at least one state.
This generates $\mathcal{B}^{(m)}$.
\item[(ii)] Attach $\mathcal{G}^{(m)}$ and $\mathcal{B}^{(m)}$
to an arbitrarily given $n$-state initial lottery.
We generate $D^{(m)}$ if $\mathcal{G}^{(m)}$ state-wise precedes $\mathcal{B}^{(m)}$
and we generate $C^{(m)}$ if $\mathcal{B}^{(m)}$ state-wise precedes $\mathcal{G}^{(m)}$.
\end{itemize}
In the transformations to $D^{(m)}$ and $C^{(m)}$ we require that
the ranking of outcomes remains unaffected and the outcomes of the resulting lotteries
remain non-negative.

Then, we state the following theorem:
\begin{theorem}\label{Th:mth-1}
Let $m\geq 2$. If $C^{(m)}$ and $D^{(m)}$ are generated by the transformations described above,
then $D^{(m)}$ is preferred (dispreferred) to $C^{(m)}$ by any DT DM
with $(-1)^{m-1}h^{(m)}\geq 0$ ($(-1)^{m-1}h^{(m)}\leq 0$).
\end{theorem}


Just like in Sections \ref{sec:Res3rd} and \ref{sec:Res4th}
we consider henceforth parsimonious subclasses of lottery pairs $C^{(m)}$ and $D^{(m)}$
that are already sufficient for signing $h^{(m)}$.
We start at the second order and proceed by simple iteration.
Let $n\geq m\geq 2$ and consider
\begin{align*}C^{(m)}_{n}=&[\ x_{1},1/n\ ;\ \ldots ;\ x_{j},1/n\ ;\\
&\quad\ \mathbf{x_{j+1},1/n}\ ;\ \ldots ;\ \mathbf{x_{j+m},1/n}\ ;\\
&\quad\qquad\ x_{j+m+1},1/n; \ \ldots\ ;\ .,1/n\ ].\end{align*}
The $m$ states in bold need to be present for any $n (\geq m)$, the remaining states are added arbitrarily
until the state probabilities sum up to 1.
The increments in the outcomes when moving to an adjacent higher state of $C^{(m)}_{n}$ are allowed to be arbitrarily non-negative and state-dependent,
as long as 
the transformations conducted below do not change the ranking of outcomes.
Taking the $(m-1)^{\mathrm{th}}$ order as a starting point,
we construct the class of lottery pairs for the $m^{\mathrm{th}}$ order in two steps:
\begin{enumerate}
\item[(i)] Concatenate a same-sized but exactly opposite $\mathcal{G}^{(m-1)}$ and $\mathcal{B}^{(m-1)}$,
with $\mathcal{G}^{(m-1)}$ state-wise preceding $\mathcal{B}^{(m-1)}$ by exactly one state.
This generates $\mathcal{G}^{(m)}$.
Next, concatenate the same $\mathcal{G}^{(m-1)}$ and $\mathcal{B}^{(m-1)}$
with $\mathcal{B}^{(m-1)}$ state-wise preceding $\mathcal{G}^{(m-1)}$ by exactly one state.
This generates $\mathcal{B}^{(m)}$.
(For $m=2$, we set $\mathcal{G}^{(2)}=[\ \delta,1/n\ ]$ and $\mathcal{B}^{(2)}=[\ -\delta,1/n\ ]$, $\delta>0$.)
\item[(ii)] 
Generate $D^{(m)}_{n}$ by attaching $\mathcal{G}^{(m)}$ to the first $(m-1)$ bold states of $C^{(m)}_{n}$
and attaching $\mathcal{B}^{(m)}$ to the last $(m-1)$ bold states of $C^{(m)}_{n}$.
\end{enumerate}
In the transformations from $C^{(m)}_{n}$ to $D^{(m)}_{n}$
we require that the ranking of outcomes remains unaffected
and the outcomes of the resulting lotteries remain non-negative.
For $m\geq 3$, we perform our sequences of squeezes and anti-squeezes at overlapping states to keep the required number of states minimal.
This yields a parsimonious approach already sufficient for signing $h^{(m)}$.

Finally, the following theorem shows that within DT the preference towards the class of lottery pairs $D^{(m)}_{n}$ and $C^{(m)}_{n}$
signs $h^{(m)}$:
\begin{theorem}\label{Th:mth-2}
Let $m\geq 2$.
If, for any $n\geq m$, a DT DM prefers (disprefers) $D^{(m)}_{n}$ to $C^{(m)}_{n}$,
then $(-1)^{m-1}h^{(m)}\geq 0$ ($(-1)^{m-1}h^{(m)}\leq 0$).
\end{theorem}




\setcounter{equation}{0}

\section{Conclusion}\label{Sec:Con}


Starting with Menezes, Geiss and Tressler (1980),
many papers have been devoted to an interpretation of the signs of the successive derivatives of the utility function
within the EU model.

In this paper we have developed a model free story
of preferences towards particular nested classes of lottery pairs
that
is appropriate to satisfy the specific
requirements of the DT model.
The story yields an intuitive interpretation, and full characterization,
of the dual counterparts of such concepts as prudence and temperance.
The direction of preference between the nested lottery pairs that are provided by the story is equivalent
to signing the 
$m^{\mathrm{th}}$ derivative of the probability weighting 
function within DT.

We have analyzed implications of our results for portfolio choice, which appear to stand in sharp contrast to
familiar implications under the EU model.
We have also shown that, where the sign of the third derivative of the utility function is connected to a savings problem,
the sign of the third derivative of the probability weighting function may be naturally linked to a self-protection problem.

Because the primal and dual stories have several aspects in common,
some of the implications of the primal story can potentially be extended to a dual world.
For instance, because it is also simple, the dual story should be as amenable to experimentation 
as the primal story. 
Another promising route for future research is that 
the dual story can serve as a building block on the basis of which it should be possible to obtain 
related interpretations for more general models of choice under risk (and ambiguity).
Indeed, now that the dual story has been told, future research can develop an interpretation to the signs of the successive derivatives
of both the utility function and the probability weighting function in the rank-dependent utility (RDU) model of Quiggin (1982)
and in prospect theory of Tversky and Kahneman (1992).
For example, one may verify that the direction of preference between
the \textit{intersection} of lottery pairs generated by primal and dual risk apportionment
signs the successive derivatives of the utility and probability weighting functions \textit{jointly}
under RDU.
This would provide a test for the null hypothesis that both $U^{(3)}\geq 0$ and $h^{(3)}\geq 0$
in the RDU environment,
as is often assumed in parametric specifications of the utility and probability weighting functions.
Furthermore, subsets of lottery pairs generated by dual risk apportionment that share the same final wealth outcomes
can be used to test the null hypothesis that (only) $h^{(3)}\geq 0$ in the RDU setting.

As such, this paper
represents 
a first step, and paves new ways,
towards the development of higher order risk attitudes in non-EU theories,
as explicitly desired by Deck and Schlesinger (2010).


\newpage

\appendix

\setcounter{equation}{0}

\section*{Appendix: Proofs}

\begin{proof} \textit{of Theorem \ref{Th:Char3-1}.}
We first note that the probabilities of generating the minimal outcome in a two-shot experiment (i.e., in two independent draws)
for the states of an arbitrarily given $n$-state lottery $A$ with equal state probabilities are
\begin{equation*}(2n-1)/n^2,\ldots,7/n^2,5/n^2,3/n^2,1/n^2,\end{equation*}
respectively.
Thus, the second dual moment of $A$ is given by
\begin{equation*}
\E{\min(A_1,A_2)}= (1/n^2) x_n+ (3/n^2) x_{n-1}+ (5/n^2) x_{n-2}+ (7/n^2) x_{n-3}
+\cdots+((2n-1)/n^2) x_1.
\end{equation*}
Hence, one readily verifies that
\begin{align*}\E{C^{(3)}}&=\E{D^{(3)}},\\
\E{\min(C^{(3)}_{1},C^{(3)}_{2})}&=\E{\min(D^{(3)}_{1},D^{(3)}_{2})}.
\end{align*}
Indeed, attaching $\mathcal{G}^{(3)}$ and $\mathcal{B}^{(3)}$ with $\mathcal{G}^{(3)}$ state-wise preceding $\mathcal{B}^{(3)}$
versus attaching $\mathcal{B}^{(3)}$ and $\mathcal{G}^{(3)}$ with $\mathcal{B}^{(3)}$ state-wise preceding $\mathcal{G}^{(3)}$
has the same (incremental) impact on the dual moments up to the second order.
Furthermore,
$F_{C^{(3)}}$ surpasses $F_{D^{(3)}}$ before crossing twice, so that
$D^{(3)}$ dominates $C^{(3)}$ in third-degree dual stochastic dominance; see e.g.,
Proposition 4.9 of Wang and Young (1998).

It then follows from a Taylor expansion of the ``dual utility premium'',
\begin{equation*}V\left[D^{(3)}\right]-V\left[C^{(3)}\right],\end{equation*}
(see Theorem 4.4 of Wang and Young (1998)) 
that when the dual moments are equal up to the second order,
$h'''\geq 0$ ($h'''\leq 0$) implies that the dual utility premium above is non-negative (non-positive).
This proves the stated result.
\end{proof}

\begin{proof} \textit{of Theorem \ref{Th:Char3-2}.}
Since $h''$ is differentiable on $(0,1)$, the third derivative of $h$ being non-negative 
is equivalent to requiring 
that
\begin{align*}\left([h(p_{+})-h(p_{0})]-[h(p_{0})-h(p_{-})]\right)-\left([h(p_{0})-h(p_{-})]-[h(p_{-})-h(p_{--})]\right)\\
=h(p_{+})-3h(p_{0})+3h(p_{-})-h(p_{--})\geq 0,
\end{align*}
for any four equidistant 
probabilities $0\leq p_{--}\leq p_{-}\leq p_{0}\leq p_{+}\leq 1$.
We will show that this condition is satisfied whenever $D^{(3)}_{n}$ is preferred to $C^{(3)}_{n}$
for any $n\geq 3$ within DT.
(The implication $h'''\leq 0$ follows similarly.)

We note that the DT evaluation of a lottery $A$ with $n$ outcomes $x_{0}=0\leq x_{1}\leq\cdots\leq x_{n}$ is given by
\begin{align*}V[A]&=\int_{0}^{\infty}x\,\mathrm{d}\left(1-\bar{h}(1-F_{A}(x))\right)
=-\sum_{i=1}^{n}x_{i}\left(\bar{h}(1-F_{A}(x_{i}))-\bar{h}(1-F_{A}(x_{i-1}))\right)\\
&=\int_{0}^{\infty}\bar{h}(1-F_{A}(x))\,\mathrm{d}x
=\sum_{i=1}^{n}\bar{h}(1-F_{A}(x_{i-1}))(x_{i}-x_{i-1}),\end{align*}
with $\bar{h}(p)=1-h(1-p)$.
Observe that $\bar{h}:[0,1]\rightarrow[0,1]$ with $\bar{h}(0)=0$, $\bar{h}(1)=1$, $\bar{h}'\geq 0$;
and that $\bar{h}''\geq 0$ is equivalent to $h''\leq 0$.
Here, $F_{A}(x_{0})=0$ by convention,
so $1-F_{A}(x_{0})=\bar{h}(1-F_{A}(x_{0}))=1$.
Hence,
\begin{align*}V[C^{(3)}_{n}]=&
\sum_{i=1}^{j}\bar{h}(1-F_{C^{(3)}_{n}}(x_{i-1}))(x_{i}-x_{i-1})+\sum_{i=j+1}^{j+4}\bar{h}(1-F_{C^{(3)}_{n}}(x_{i-1}))(x_{i}-x_{i-1})\\
&+\sum_{i=j+5}^{n}\bar{h}(1-F_{C^{(3)}_{n}}(x_{i-1}))(x_{i}-x_{i-1})\\
=&:\Sigma_{1}^{C^{(3)}_{n}}+\Sigma_{2}^{C^{(3)}_{n}}+\Sigma_{3}^{C^{(3)}_{n}},
\end{align*}
with $j$ corresponding to the state preceding the three states in bold that are always present in $C^{(3)}_{n}$, and similarly for $D^{(3)}_{n}$.
Notice that
\begin{equation*}\Sigma_{1}^{C^{(3)}_{n}}=\Sigma_{1}^{D^{(3)}_{n}},\qquad \Sigma_{3}^{C^{(3)}_{n}}=\Sigma_{3}^{D^{(3)}_{n}}.\end{equation*}
Furthermore,
\begin{align*}\Sigma_{2}^{C^{(3)}_{n}}=\ &(x_{j+1}-x_{j})\bar{h}(1-F(x_{j}))+(x_{j+2}-x_{j+1})\bar{h}(1-F(x_{j+1}))\\
&+(x_{j+3}-x_{j+2})\bar{h}(1-F(x_{j+2}))+(x_{j+4}-x_{j+3})\bar{h}(1-F(x_{j+3})),\end{align*}
and
\begin{align*}\Sigma_{2}^{D^{(3)}_{n}}=\ &(x_{j+1}-x_{j}+1/M)\bar{h}(1-F(x_{j}))+(x_{j+2}-x_{j+1}-3/M)\bar{h}(1-F(x_{j+1}))\\
&+(x_{j+3}-x_{j+2}+3/M)\bar{h}(1-F(x_{j+2}))+(x_{j+4}-x_{j+3}-1/M)\bar{h}(1-F(x_{j+3})),\end{align*}
suppressing the indices $C^{(3)}_{n}$ and $D^{(3)}_{n}$ for convenience.
Define, for given $n\geq 3$, the probability $q$, $0\leq q\leq 1-3/n$, such that
\begin{align*}F(x_{j})=q,\qquad F(x_{j+1})=q+1/n,\qquad F(x_{j+2})=q+2/n,\\
F(x_{j+3})=q+3/n,\qquad F(x_{j+4})=q+4/n.\end{align*}
Then,
\begin{align*}\Sigma_{2}^{C^{(3)}_{n}}=\ &(x_{j+1}-x_{j})\bar{h}(1-q)+(x_{j+2}-x_{j+1})\bar{h}(1-q-1/n)\\
&+(x_{j+3}-x_{j+2})\bar{h}(1-q-2/n)+(x_{j+4}-x_{j+3})\bar{h}(1-q-3/n),\end{align*}
and
\begin{align*}\Sigma_{2}^{D^{(3)}_{n}}=\ &(x_{j+1}-x_{j}+1/M)\bar{h}(1-q)+(x_{j+2}-x_{j+1}-3/M)\bar{h}(1-q-1/n)\\
&+(x_{j+3}-x_{j+2}+3/M)\bar{h}(1-q-2/n)+(x_{j+4}-x_{j+3}-1/M)\bar{h}(1-q-3/n).\end{align*}
Upon defining $p_{--}, p_{-}, p_{0}, p_{+}$ such that
\begin{align*}3/n\leq 1-q=p_{+}\leq 1,\qquad 1-q-1/n=p_{0},\\1-q-2/n=p_{-},\qquad
1-q-3/n=p_{--},
\end{align*}
it then follows from the arbitrariness of $n\geq 3$, hence the arbitrariness of $0\leq q\leq 1-3/n$, that
\begin{align*}\Sigma_{2}^{D^{(3)}_{n}}-\Sigma_{2}^{C^{(3)}_{n}}=
(1/M)\bar{h}(p_{+})-(3/M)\bar{h}(p_{0})+(3/M)\bar{h}(p_{-})-(1/M)\bar{h}(p_{--})\geq 0,\end{align*}
for any four equidistant 
probabilities $0\leq p_{--}\leq p_{-}\leq p_{0}\leq p_{+}\leq 1$,
whenever $D^{(3)}_{n}$ is preferred to $C^{(3)}_{n}$ for any $n\geq 3$.
We note that $\Sigma_{2}^{D^{(3)}_{n}}-\Sigma_{2}^{C^{(3)}_{n}}$
is the DT analog of the utility premium in the EU model.
Finally, observe that $\bar{h}'''\geq 0 $ is equivalent to $h'''\geq 0$.
This proves the stated result.
\end{proof}

\begin{proof} \textit{of Theorem \ref{Th:Char4-1}.}
By analogy to the proof of Theorem \ref{Th:Char3-1},
we first note that the probabilities of generating the minimal outcome in a three-shot experiment
(that is, in three independent draws)
for the states
of an arbitrarily given $n$-state lottery $A$ with equal state probabilities are
\begin{equation*}
(3(n-1)n+1)/n^3,\ldots,37/n^3,19/n^3,7/n^3,1/n^3,
\end{equation*}
respectively.
Thus, the third dual moment of $A$ is given by
\begin{align*}
\E{\min(A_1,A_2,A_3)}=\ & (1/n^3) x_n+ (7/n^3) x_{n-1}+ (19/n^3) x_{n-2}+ (37/n^3) x_{n-3}\qquad\qquad\qquad\qquad\qquad
\\&+\cdots+((3(n-1)n+1)/n^3) x_1.
\end{align*}
Hence, one may verify that not only the first two dual moments of $C^{(4)}$ and $D^{(4)}$ are equal, but also the third:
\begin{align*}
\E{\min(C^{(4)}_{1},C^{(4)}_{2},C^{(4)}_{3})}&=\E{\min(D^{(4)}_{1},D^{(4)}_{2},D^{(4)}_{3})}.
\end{align*}
Indeed, attaching $\mathcal{G}^{(4)}$ and $\mathcal{B}^{(4)}$ with $\mathcal{G}^{(4)}$ state-wise preceding $\mathcal{B}^{(4)}$
versus attaching $\mathcal{B}^{(4)}$ and $\mathcal{G}^{(4)}$ with $\mathcal{B}^{(4)}$ state-wise preceding $\mathcal{G}^{(4)}$
has the same (incremental) impact on the dual moments up to the third order.
The proof then follows by similar arguments as the proof of Theorem \ref{Th:Char3-1}.
The details are suppressed to save space.
They are contained in the extended online version.
\end{proof}

\begin{proof} \textit{of Theorem \ref{Th:Char4-2}.}
The proof follows by similar arguments as the proof of Theorem \ref{Th:Char3-2}.
The details are suppressed to save space.
They are contained in the extended online version.
\end{proof}

\begin{proof} \textit{of Theorems \ref{Th:mth-1} and \ref{Th:mth-2}.}
The proofs follow by similar arguments as the proofs of Theorems \ref{Th:Char3-1} and \ref{Th:Char4-1}
and Theorems \ref{Th:Char3-2} and \ref{Th:Char4-2}, respectively.
See the extended online version for some further details.
%
%
%
\end{proof}

\baselineskip 0.16 cm

\begin{spacing}{0.7}

\end{spacing}

\end{document}